\documentclass[a4paper,11pt]{article}

\pdfoutput=1 

\usepackage{jheppub} 

\usepackage[T1]{fontenc} 
 
 \usepackage{braket}

 \usepackage{tikz}
 \usepackage{pgfplots}
 \usepgfplotslibrary{fillbetween}

\DeclareMathOperator{\E}{\mathsf{E}} 
\newcommand{\R}[0]{\mathbb{R}} 
\renewcommand{\O}{\mathcal{O}}
\newcommand{\norm}[1]{\left\lVert#1\right\rVert}

\DeclareMathOperator*{\argmin}{arg\,min}

\title{Prospects and challenges of quantum finance}

\author[a,b]{Adam Bouland,}
\author[a,c,d]{Wim van Dam,}
\author[a]{Hamed Joorati,}
\author[a,e]{Iordanis Kerenidis,}
\author[a]{Anupam Prakash}

\affiliation[a]{QC Ware Corp, 550 Hamilton Ave, Palo Alto, CA, USA}
\affiliation[b]{Department of EECS, University of California, Berkeley, USA}
\affiliation[c]{Department of Computer Science, University of California, Santa Barbara, USA}
\affiliation[d]{Department of Physics, University of California, Santa Barbara, USA}
\affiliation[e]{CNRS, University of Paris Diderot, Paris, France}

\emailAdd{\{adam.bouland, wim.van.dam, hamed.joorati, iordanis.kerenidis, anupam.prakash\}@qcware.com}

\abstract{
Quantum computers are expected to have substantial impact on the finance industry, as they will be able to solve certain problems considerably faster than the best known classical algorithms.
In this article we describe such potential applications of quantum computing to finance, starting with the state-of-the-art and focusing in particular on recent works by the QC Ware team. 
We consider quantum speedups for Monte Carlo methods, portfolio optimization, and machine learning. 
For each application we describe the extent of quantum speedup possible and estimate the quantum resources required to achieve a practical speedup.
The near-term relevance of these quantum finance algorithms varies widely across applications --- some of them are heuristic algorithms designed to be amenable to near-term prototype quantum computers, while others are proven speedups which require larger-scale quantum computers to implement.
We also describe powerful ways to bring these speedups closer to experimental feasibility --- in particular describing lower depth algorithms for Monte Carlo methods and quantum machine learning, as well as quantum annealing heuristics for portfolio optimization. 
This article is targeted at financial professionals and no particular background in quantum computation is assumed. 
}

\pgfplotsset{compat=1.16}

\begin{document} 
\maketitle
\flushbottom
\newpage
\section{Quantum computing in finance}

Quantum computers are expected to have substantial impact on industry, as they will be able to solve certain problems considerably faster than the best known classical algorithms. However, quantum computers are not merely faster processors that one can blindly use instead of classical computers to speed up all problems. They are a fundamentally different way of performing computation that has the power, when one applies the right quantum algorithms, to provide considerable speedups for  \emph{specific} computational tasks. Therefore there is an important algorithmic question to address: does my application admit any quantum speedup? This is a necessary precursor to establishing a timeline for commercial relevance. In case one can prove that a quantum speedup exists for a problem of interest, and there are indeed many such examples, there is the question of when this speedup will become practical. To date, quantum hardware is not advanced enough for solving any problem of practical relevance faster than classical computers. Nevertheless, in 2019 we saw the first demonstration of a quantum speedup, a milestone result in itself, for a contrived mathematical problem designed to play to the strengths of the prototype quantum device \cite{arute2019quantum}. To determine when the quantum speedup will become relevant for a particular domain requires a deep understanding of the quantum algorithms themselves, of the quantum resources required to run the algorithms, the extent to which those resources can be reduced by redesigning the algorithms, as well as estimating the timeline of the development of quantum hardware. 

Modern computational finance offers a host of challenging computational problems with which to explore the application of quantum computers. Quantitative analysts face a variety of challenges spanning optimization and machine learning. Common problems include: evaluating risk and pricing exotic derivatives using Monte Carlo methods, modeling financial markets with stochastic differential equations, optimizing portfolio allocations, and identifying market trends with machine learning techniques. Solving these problems often requires sophisticated computational methods, and in some cases substantial computational resources as well. A natural question is therefore: how will the arrival of quantum computers affect computational finance? For which of these computational problems will quantum computers offer a speedup? And given the nascent state of quantum hardware, how soon can these speedups become commercially relevant?

Here, we review quantum algorithms for computational finance for the quantitative analyst. We provide an in-depth look into several applications of quantum computing to finance, focusing on quantum speedups for Monte Carlo methods (Section \ref{sec:montecarlo}), portfolio optimization (Section \ref{sec:portfolioopt}), and machine learning (Section \ref{sec:ML}). For each application we describe the potential quantum speedup over classical methods (if known), and the quantum hardware resources required to run it at commercial scale. Moreover, we describe our ongoing work and future research directions to lower the resource requirements of these algorithms, and therefore potentially bring forward their commercial relevance. This complements recent surveys of Orus et al.\ \cite{orus2019quantum} and Egger et al.\ \cite{egger2020survey} who provided a higher level overview of the subject.

Broadly speaking, quantum algorithms for finance fall into two categories: those algorithms which are feasible in near-term quantum computing hardware of the next several years, and those which will require substantially more advanced quantum hardware (i.e. fault-tolerant quantum computers) to run.
This distinction arises because current prototype quantum computers are extremely noisy.
For context, the noise rate per fundamental quantum operation --- analogous to a single NAND gate in a classical CPU --- is roughly $0.6\%$ \cite{arute2019quantum}. In contrast the physical noise rate of modern classical computers is negligible, as the noise from CPU error is dwarfed by the storage error rate of RAM. 
Therefore current quantum devices cannot perform more than $10^2 - 10^3$ simple operations before devolving into noise.
In the long run it is well known that quantum error correction techniques will allow one to mitigate these errors by redundantly encoding the information in the physical system, much in the same way that classical error correcting codes trade redundancy for decreased error rates (see e.g. \cite{gottesman1997stabilizer,calderbank1997quantum,lidar2013quantum}). 
This decreases the effective error rate, but increases the number of qubits needed for the redundant encoding, and also decreases the effective clock speed of the device.
This additional qubit overhead is quite steep\footnote{Current estimates are ~$10^3$-$10^4$ physical qubits per logical qubit \cite{fowler2018low}, but this estimate is very sensitive to assumptions about the quantum error rate of devices with a large number of qubits. } and is out of reach of current devices with limited number of qubits.
The effective clock speed may also be slowed by several orders of magnitude \cite{gidney2019efficient}.
As a result, in the next few years prototype quantum computers will be in the ``Noisy Intermediate-Scale Quantum'' (NISQ) regime \cite{preskill2018quantum}, in which only short, non-error-corrected computations can be carried out.
Certain quantum algorithms have been designed to operate in this regime (e.g. \cite{farhi2014quantum}), while others have been a priori designed with fully fault-tolerant devices in mind (e.g. \cite{shor1999polynomial,gidney2019factor}).

We note that quantum solutions for finance which have been specifically designed for NISQ implementations do not have convincing evidence for quantum speedups, now or in the long run.
This is for several reasons.
First, most of these quantum algorithms are \emph{heuristic} in nature --- they do not have proven performance guarantees (or if they do, they are surpassed by classical algorithms), but neither do many classical heuristics used in practice.
It is extremely difficult to compare the performance of heuristic algorithms, as oftentimes heuristics will perform better on certain problem instances and worse on others.
To this end we advocate the creation of a public repository of basic classical heuristics for finance problems such as portfolio optimization, as well as toy instances of those problems, to better facilitate the development of quantum heuristics and provide fair comparisons. 
Second, and perhaps more fundamentally, in many cases it is unclear if a quantum speedup will even exist for these heuristics. 
For example, it has recently been suggested that simple classical heuristics might outperform analogous quantum heuristics at optimization problems \cite{hastings2019classical} --- see Section \ref{sec:portfolio_combinatorial_futuresprospects} for further discussion. 
It is also a natural question whether or not some of the ``quantumness'' can be removed from these NISQ heuristics, and if they can be replaced by ``quantum-inspired'' classical heuristics. 
Finally, for many of these proposed NISQ heuristics, there does not yet exist quantum hardware of sufficient scale and quality in order to run the algorithm on a large enough instance of a problem to be competitive with a classical solver.
For example, in the case of portfolio optimization, the largest instances run on hardware have been relatively small due to overheads of embedding the problem into quantum hardware.
These instance sizes are trivial for classical solvers, so such small proof-of-concept implementations do not shed light onto potential speedups for larger sized problems.
This calls for a more detailed examination of the impact of these NISQ heuristics from a classical and quantum algorithmic perspective. More importantly, the power of heuristics most often comes from deep domain specific knowledge and thus the collaboration between quantum algorithmic and finance experts is essential in order to be able to increase the impact of quantum heuristics and harness the computational potential of the NISQ machines. 

On the other hand, a number of powerful quantum algorithms for finance have been proven to provide considerable asymptotic speedups over classical methods. This is true for some of the main computational challenges in finance, including for Monte Carlo methods \cite{montanaro2015quantum}, combinatorial portfolio optimization \cite{montanaro2020quantum}, convex portfolio optimization \cite{rebentrostlloyd2018quantumcf,kerenidis2019quantum}, and general machine learning \cite{KP16}. However, out-of-the-box implementations of such algorithms with provable guarantees all seem to require a priori full-scale fault-tolerant quantum computing devices.
One of the most pressing challenges, then, is to redesign these algorithms in order to both considerably reduce the quantum hardware requirements and at the same time keep their provable impact. 

To this end, in addition to a review of the state-of-the-art, we also describe several methods we have developed at QC Ware for designing or redesigning quantum algorithms with reduced resource requirements while maintaining performance guarantees. We particularly focus on improvements to the algorithms themselves, resulting to resource reductions of many orders of magnitude, which go beyond simple optimizations of the circuit implementation of the algorithms (which in their turn can also offer further reductions). We briefly provide the main ideas here and provide more detailed descriptions in the following sections.

\subsubsection*{Monte Carlo methods}

Let us start with quantum Monte Carlo methods, one of the most prominent applications of quantum algorithms in finance. It has been shown that full-scale fault-tolerant quantum computers offer a significant speedup, by drastically reducing the number of samples needed compared to classical Monte Carlo methods \cite{heinrich2001quantum,montanaro2015quantum}. This has the potential to provide more accurate or more stable pricing, or could allow for close to real-time pricing of derivatives which would otherwise require slower Monte Carlo evaluations --- potentially providing a strong competitive advantage for users of this technology.
At the same time, first estimations seem to show that this quadratic speedup achieved for standard quantum Monte Carlo methods may be limited in its applicability to rather difficult Monte Carlo problems and may require very high quality qubits to execute. 
This would require high overheads in the number of physical qubits and effective quantum clock speed due to the need to implement rigorous quantum error correction (see Section \ref{sec:MCresourceestimates}).
Therefore these quantum algorithms cannot be used ``out of the box'' for the foreseeable future, and even then may be limited in their applicability. This is also true for more recent proposals \cite{suzuki2020amplitude,aaronson2020quantum} where the depth of the quantum circuit remains asymptotically the same, even though a small reduction is achieved.
A natural question to ask is whether one can design hybrid NISQier algorithms that will realize a smooth continuum between fully quantum and classical Monte Carlo algorithmically, using circuits of much lower depth and retaining a provable speedup. Recent work of Kerenidis and Prakash \cite{KPpatent}, building on prior works of Burchard \cite{burchard2019lower} and Suzuki et al. \cite{suzuki2020amplitude}, has shown the existence of such algorithms, bringing the prospect of quantum speedups for Monte Carlo methods closer to realization.
This quantum algorithm starts providing a significant speedup over the classical Monte Carlo methods with much lesser quality qubits than the standard quantum algorithm, and its performance scaling is optimal (see Section \ref{sec:MCimprovements} for details).
We describe in Section \ref{sec:MCresourceestimates} how this improvement opens the door to quantum speedups at higher effective error rates in the quantum device, which could potentially bring the timeline of this algorithm's relevance forward.

\subsubsection*{Portfolio Optimization}

Another important application of quantum algorithms in finance is for portfolio optimization. We will present separately the case where the portfolio optimization can be cast as a convex program, where we only consider linear budget constraints, for example for long-only portfolios; and the more general case where the optimization can also include integer constraints, for example when a limit on the number of assets to invest is set. For the first case, we start by describing quantum algorithms with provable speedups for certain cases, based on quantum linear system solvers \cite{rebentrostlloyd2018quantumcf,kerenidis2019quantum}. The main limitation of these algorithms is their dependence on quantum linear system solvers \cite{HHL09,CGJ18, GLSW18} that require large depth circuits and remain beyond the reach of
near and intermediate term devices. One of the main challenges for bringing quantum portfolio optimization closer to reality is finding ways to redesign quantum linear systems solvers, and quantum linear algebra more generally, with reduced hardware requirements. We provide more details about these algorithms and {preliminary ideas for bringing quantum linear algebra closer to the NISQ era} in the following sections.
For the second case, we look at portfolio optimization as a combinatorial optimization problem, for which we describe various NISQ heuristics both for quantum annealers and for quantum circuits. We also include a discussion of some specialized tools we have developed to improve the performance of standard quantum optimization heuristics, such as anneal offsets \cite{adame2020inhomogeneous}.

\subsubsection*{Machine Learning in Finance}

In the third part, we turn our attention to Machine Learning, which has become ubiquitous in recent times, with the growth in the types and size of data sets and the development of high-performance machine learning algorithms for a wide range of data analysis tasks. Quantum machine learning algorithms with potentially significant polynomial speedups have been proposed for many fundamental machine learning tasks \cite{LMR13, LMR14, KP16, qmeans}. Once more, ``out-of-the-box'' quantum machine learning algorithms require full scale quantum computers and moreover, assume the availability of a quantum random access memory (QRAM), which is a quantum analogue of the classical Random Access Memory (RAM) and may be difficult to construct. Hence, before being able to bring quantum machine learning closer to the NISQ era, one would need efficient ways to load classical data into quantum states and perform fast computations with them. We recently developed such efficient ways for encoding classical data into quantum states using parametrized quantum circuits, called data loaders \cite{patent2}, which can be very fast and can be used by a quantum algorithm, for example to estimate the distance between data points in supervised and unsupervised similarity learning, or to perform simple linear algebra procedures for training neural networks. This work significantly pushes quantum machine learning algorithms closer to the NISQ era than previously thought. Another interesting approach in quantum machine learning is to design quantum analogues of neural networks and heuristic methods for training such quantum circuits to perform classification or other learning tasks. While such quantum neural networks may eventually provide better performance than classical networks, it is very difficult to provide any convincing evidence due to the absence of theoretical analysis or large-scale quantum simulations.
We discuss some of the advances in quantum machine learning briefly below with a view towards applications to finance. 

It should be clear by now that quantum computing has the potential to impact considerably the finance sector, provably with the advent of full-scale quantum computers and possibly in the NISQ era as well, both through the NISQ-ification of powerful quantum algorithms and the better design and tuning of quantum heuristics. Finding these first quantum finance applications is a formidable challenge and it is certainly the time for both quantum algorithmic and finance experts to be put together to the task. 

\section{Monte Carlo methods for derivatives pricing and risk management}
\label{sec:montecarlo}

One of the most prominent applications of quantum algorithms in finance is the improvement of Monte Carlo methods.
Monte Carlo methods are a broad class of algorithms based on randomized sampling.
In most applications of Monte Carlo methods in finance, one starts with a stochastic model $S$ of the market --- for instance, of the price trajectories of various securities evolving by a random walk.
The goal of the algorithm is to estimate some particular function $f$ of the price trajectories.
For instance, in Value at Risk (VaR) calculations, one wishes to estimate the amount of value a portfolio could lose over a particular time horizon excluding some fixed fraction of outlier events.
Another application is in the pricing of financial derivatives, which can be complicated functions of market prices. 
Here the price $V$ of the derivative is the expected value of $f$, averaged over future market conditions.
In both cases, to estimate the desired quantity using Monte Carlo method, one simply simulates the stochastic model $S$ starting with many random seeds and takes the empirical average of the desired quantity $f$. 
For instance, in derivative pricing $S$ could represent one's stochastic model of the market (e.g. a lognormal random walk of asset prices), $f$ could represent the payoff of the derivative under those market conditions, and therefore the quantity to estimate $V$ will represent the (risk-neutral) price one should pay for the derivative.

More formally, suppose that one has a stochastic market model $S:\{0,1\}^r \rightarrow \{0,1\}^n$ which takes as input an $r$-bit random seed, and outputs an $n$-bit description of the market price trajectories $x$.
Additionally, one has a payoff function $f:\{0,1\}^n \rightarrow \mathbb{R}$, and one wishes to estimate 
\begin{align*}
V& \mathrel{\mathop:}=\displaystyle\E_{x\sim S} f(x).
\end{align*}
For special choices of models $S$ and payoff functions $f$ there exist analytical formulae for $V$ --- for example the Black-Scholes formula for the pricing of European call options for assets evolving under geometric Brownian motion.
However in general no analytical formula exists, even for fairly simple option payoff functions, and therefore instead $V$ is estimated by Monte Carlo sampling.
That is, one takes say $N$ samples $x_1\ldots x_N$ from the model $S$ (by choosing $N$ random seeds), and uses the empirical average of the samples
\begin{align*}
\tilde{V} & \mathrel{\mathop:}= \frac{1}{N}\left(f(x_1)+\ldots+f(x_N)\right)
\end{align*}
as the estimate for $V$.
It is well known that the convergence of this algorithm is quadratic in the number of samples --- that is to achieve accuracy $\epsilon$ in the value of $V$ one needs to take roughly $N=O\left(1/\epsilon^2\right)$ samples. 
More precisely if $\sigma$ is the standard deviation of the random variable $f(x)$ for $x$ drawn from $S$, then we have that
\begin{align*}
\Pr[|V-\tilde{V}|<\epsilon] & \geq \frac{\sigma^2}{N\epsilon^2}
\end{align*}
Therefore the $99\%$ confidence interval $\epsilon$ scales as
\begin{align*}
\epsilon=O\left(\frac{\sigma}{\sqrt{N}}\right)
\end{align*}
with the number of samples $N$.
To put this another way, to achieve $99\%$ confidence that one has $\epsilon$-approximated $V$ requires $N=O\left(\sigma^2/\epsilon^2\right)$ samples.

This quadratic dependence on epsilon can be quite expensive in practice; getting an accuracy of $10^{-3}$ or $10^{-5}$ in derivative pricing values can require millions to billions of Monte Carlo samples. 
The practical cost of this can vary widely between applications --- pricing simple derivatives on only a few assets and to not-so-high accuracy can take few seconds, while performing high accuracy Value at Risk calculations for complex structured products can take substantially longer, even on large clusters.

\subsection{Quantum speedups for Monte Carlo}

Quantum computing offers a significant speedup for Monte Carlo methods, quadratically improving the dependence on $\sigma$ and $\epsilon$.
In particular, Montanaro showed that there is a quantum algorithm which estimates the mean $V$ to accuracy $\epsilon$ using at most 
\begin{align*}
N=\tilde{\O}\left(\frac{\sigma}{\epsilon}\right)
\end{align*}
queries to the stochastic model $S$ and payoff function $f$ \cite{montanaro2015quantum}.
This builds on a long history of works in quantum algorithms: first, quantum algorithms were developed to $\epsilon$-approximate the mean of a bounded function $f:X\rightarrow [0,1]$ under the uniform distribution on the domain $X$ in $ \O(1/\epsilon)$ time \cite{abrams1999fast,aharonov1999quantum,grover1998framework,heinrich2001quantum,brassard2011optimal}.
This algorithm was subsequently generalized to the case of arbitrary distributions over $X$ when there is a bound on the variance of $f(x)$ \cite{montanaro2015quantum}, or when one has further information about the distribution such as upper and lower bounds on the mean \cite{li2018quantum,hamoudi2018quantum}.
At their core, these algorithms make use of an algorithmic primitive known as quantum amplitude estimation \cite{brassard2002quantum}, which is closely related to the quantum approximate counting algorithm \cite{brassard1998quantum}.
This quantum algorithm drastically reduces the number of calls one needs to make to the stochastic model $S$ and payoff function $f$ --- in particular, achieving accuracy of say $\epsilon=10^{-3}$ now merely requires $N\approx  \O(10^3)$ calls to $S$, a significant improvement over classical Monte Carlo sampling.
In the long run, with fully scalable fault-tolerant quantum computers, this has the potential to allow for close to real-time pricing of derivatives which would otherwise require slower Monte Carlo evaluations --- potentially providing a strong competitive advantage for users of this technology. Of course when evaluating this quantum speedup, one should also compare to classical methods to accelerate Monte Carlo, such as importance sampling and other variance reduction techniques.
We note it might be possible to incorporate these techniques into quantum algorithms for Monte Carlo as well, and this is an interesting question for future research.

This has generated a huge amount of interest in quantum algorithms for Monte Carlo from academics, quantum startups, and the financial sector.
For example, several works have explored the concrete application of the quantum algorithm to particular financial applications, such as pricing simple call options and bonds, and Value at Risk calculations \cite{rebentrost2018quantum, woerner2019quantum, egger2019credit}.
Many of these works have focused on optimizing the implementation of the Monte Carlo algorithm to those particular applications \cite{stamatopoulos2019option,ramos2019quantum,miyamoto2019reduction,vazquez2020efficient,kaneko2020quantum}.
There have also been toy experimental demonstrations of simple Monte Carlo calculations in quantum hardware \cite{woerner2019quantum,stamatopoulos2019option}.

More specifically, Rebentrost, Gupt and Bromley \cite{rebentrost2018quantum} applied the algorithm to pricing European and Asian Options.
Woerner and Egger \cite{woerner2019quantum} applied the algorithm to Value at Risk and Conditional Value at Risk calculations, and implemented this on highly simplified examples (e.g. fixed rate bonds) in IBM hardware and in simulation. 
Egger et al.\ \cite{egger2019credit} applied the algorithm to estimate the economic capital requirements, and also provided estimates of the quantum resources needed to run the algorithm at commercial scale.
Stamatopoulos \emph{et al. } \cite{stamatopoulos2019option} applied the algorithm to more complicated options such as multi-asset and path dependent options, and also ran a highly simplified proof of concept on IBM hardware.
Ramos et al.\ \cite{ramos2019quantum} described an alternative way of encoding the algorithm using more qubits but with lower circuit depth --- namely using unary encoding of the financial variables.
In the other direction, Miyamoto and Shiahara \cite{miyamoto2019reduction} described an encoding which would use fewer qubits but requires higher circuit depth, using pseudorandom number generators, and Kaneko et al.\ \cite{kaneko2020quantum} describes this encoding's effects on gate counts for the local volatility model.
Finally, Vazquez and Woerner \cite{vazirani2013approximation} described techniques for reducing the overheads of state preparation for the algorithm (that is, reducing the costs of the quantum implementation of the stochastic market model $S$).

\subsection{Limitations of the standard quantum algorithm}
\label{sec:MClimitations}

We emphasize that to date this algorithm has only only been performed on systems with a handful of qubits, and has not yet demonstrated an advantage over classical methods due to the small system sizes considered.
For example, Stamatopoulos et al.\ \cite{stamatopoulos2019option} and Woerner and Egger \cite{woerner2019quantum} ran their proof-of-concept algorithms on systems of 3 and 4 qubits, respectively (that is, those implementations could trivially be simulated on a laptop by multiplying $16\times 16$ matrices).
This is because the algorithm has very high hardware requirements which prevent it from being run at scale on current experimental quantum computers. We will describe these limitations in detail below.

First, the stochastic model $S$ and the payoff function $f$ must be run \emph{in superposition} by the quantum computer.
That is to say, one cannot simply take $ \O(1/\epsilon)$ samples from $S$ and compute $f$ with existing classical computing hardware, and use a quantum computer to extract the answer.
Instead, one has to run the entire classical stochastic model $S$ and payoff function $f$ in ``quantum-native'' hardware on the quantum computer itself.
In current quantum computing hardware, this is not yet possible except for very simple payoff functions and market models, such as those involved in pricing a bond with a fixed return or a very simple call option \cite{woerner2019quantum,stamatopoulos2019option}.
This is due to the high amount of noise present in experimental quantum computers.
As previously mentioned, the noise rate per fundamental two-qubit gate operation --- equivalent to applying a single NAND operation in the classical context --- in state of the art quantum devices is of the order $0.6\%$ \cite{arute2019quantum}. 
Thus, one can only perform $\sim10^2$-$10^3$ operations on state of the art quantum devices before the device devolves into noise.
For this reason, there has been much work to try to optimize the implementations of these market model and payoff functions to compile them to quantum hardware using a few quantum gates and as few qubits as possible \cite{stamatopoulos2019option,ramos2019quantum,miyamoto2019reduction,vazquez2020efficient}.
This sometimes requires optimizations which will not scale well asymptotically, but do improve performance with small-scale and highly noisy devices \cite{ramos2019quantum}.

Second, and perhaps more fundamentally, these quantum implementations of the market model and payoff function must be performed many times \emph{in series} on the quantum computer.
That is to say, one does not simply apply $S$ and $f$ once in superposition, stop the quantum computation, and perform classical postprocessing. Rather, in the quantum algorithm one applies $S$ and $f$ (in superposition) on a quantum computer $\tilde{\O}(\sigma/\epsilon)$ times in series before concluding the quantum portion of the computation.
Even with modest values of $\epsilon$ this requires the ability not only to compute $S,f$ in superposition, but moreover the ability to do so thousands of times in series without error.

Taken together, these two limitations suggest that to run this particular quantum algorithm at commercial scale, the effective quantum error rate of the hardware would likely need to be very small, in order to ensure that no errors\footnote{Of course there is the possibility that the algorithm could work even if a small number of errors occur in the process of the computation, but our evidence so far indicates that the algorithm is fairly sensitive to noise (e.g. \cite{regev2008impossibility,campbell2019applying}). } occur in the serial applications of $S$ and $f$ (along with other quantum operations between these applications). 
While it is well-known that one can use quantum error correction to efficiently lower the effective error rates of quantum devices, this comes at a high cost in terms of the number of qubits needed and in the time required to run the algorithm.
These limitations of the standard quantum algorithm and the still nascent state of quantum hardware are why the existing experimental implementations have been only for toy problems, such as pricing a bond, and not for more complicated payoff functions with commercial value.

In the next section we examine work being done to make the algorithm more near-term feasible than it currently is. We will then also provide resource estimates for the standard and these improved algorithms.

\subsection{Reducing the resources required for quantum Monte Carlo speedups}
\label{sec:MCimprovements}

As we discussed, the major limitation preventing the early commercial application of quantum algorithms to Monte Carlo is the \emph{circuit depth} of the algorithm. 
The algorithm requires running a large number of applications of the stochastic market model $S$ and payoff function $f$ in series, followed by a quantum Fourier transform.
This necessitates very low error rates in the quantum hardware, which are currently unavailable.
The major open problem is to determine if, and to what extent, this depth requirement can be relaxed.
For simplicity, in this section we focus on the special case where the payoff function is bounded, though similar statements hold for the more general case.

To date there have been two main approaches proposed to lower the depth of the algorithm.
The first approach has been to eliminate the quantum Fourier transform (QFT) from the algorithm \cite{suzuki2020amplitude,aaronson2020quantum,grinko2019iterative}.
This is the final step of the algorithm which occurs after the controlled serial applications of the market model $S$ and payoff function $f$. 
Asymptotically speaking, the elimination of the QFT is negligible in the overall circuit depth of the algorithm.
This is because, to approximately count to error $\epsilon$ requires $ \O(1/\epsilon)$ applications of market model $S$ and payoff function $f$ in series, but the corresponding quantum Fourier transform only occurs over a ``count'' register of $ \O(\log(1/\epsilon))$ qubits, and on those only has depth $ \O(\log\log(1/\epsilon))$ \cite{cleve2000fast}. 
However practically speaking this elimination of the QFT could have benefits.
This is because it eliminates the need to apply the market model $S$ and payoff function $f$ as controlled operations (which would generally be done gate by gate), and therefore improves the locality of the gates in the algorithm.
Since normally $k+1$-local gates are implemented using a large constant number of $k$-local gates, this improves the circuit depth by a significant (but asymptotically ignored) constant factor.

The second, and more critical question, is whether or not one can use fewer than $ \O(1/\epsilon)$ serial applications of $S$ and $f$ in the quantum algorithm. 
In the standard quantum Monte Carlo algorithm, one applies $ \O(1/\epsilon)$ serial applications of $S$ and $f$, but one only needs to perform this algorithm \emph{once} to compute the desired expectation value with high probability.
In contrast, classical Monte Carlo requires $ \O(1/\epsilon^2)$ applications of $S$ and $f$, but these applications can be performed in parallel.
One question that arises is how much one can learn with parallel applications of the quantum counting algorithm with \emph{less} than $ \O(1/\epsilon)$ depth. 
In other words, can we trade reduced circuit depth for an increased number of parallel runs of the algorithm?
If so then one could efficiently parallelize the standard quantum algorithm, allowing us to recover the quantum speedup at lower depth.

We know that there are limitations to these approaches, because (generically speaking) one \emph{cannot} reduce the circuit depth of the algorithm while maintaining the \emph{entire} speedup.
In particular, a recent work of Burchard \cite{burchard2019lower}, extending prior work of Zalka\footnote{In particular, Zalka showed that Grover search only admits trivial parallelization: to find an item in a list of $N$ items with $p$ parallel runs of a quantum algorithm requires $\Omega\left(\sqrt{\frac{N}{p}}\right)$ queries to the search oracle. So the best one can do is simply divide the domain into $p$ separate regions and run Grover search separately on each.  Burchard \cite{burchard2019lower} extended this sort of argument to the regime of approximate counting rather than search, using techniques from \cite{jeffery2017optimal}.} \cite{zalka1999grover}, showed there is a tradeoff between the depth of the quantum algorithm and the number of times it needs to be executed.
Essentially, Burchard showed that any quantum algorithm making $p$ parallel calls to the market model $S$ and bounded payoff functions $f$, and which calls them at most $D$ times in series must have that $pD^2\geq\Omega(1/\epsilon^2)$, where $\epsilon$ is the width of the confidence interval obtained by the algorithm.
(In contrast, classically one has that $pD\geq\Omega(1/\epsilon^2)$). 
The total amount of computational work needed to run the algorithm is roughly $pD$. 
Therefore, as one decreases the depth of the parallel iterates, the quantum speedup of this method must gradually disappear.
This is because of a blowup of the number of parallel runs of the quantum sampling algorithm required, which approaches the classical algorithm's parallelization of depth $D=1$ and $p=1/\epsilon^2$ parallel runs.
Essentially depth-$D$ parallel runs of the algorithm can at most achieve a $D$-fold speedup over classical Monte Carlo\footnote{Or to put it another way, $p$-fold parallelization of the quantum algorithm can only speed it up by a factor of $ \O(\sqrt{p})$.}.

While these bounds say that one cannot recover the \emph{entire} speedup at lower circuit depths, a natural question is whether or not one can recover the \emph{partial} speedup that is available at lower depth. 
In other words, can one realize a smooth continuum between quantum and classical Monte Carlo algorithmically, using circuits of lower depth, and matching the Zalka/Burchard lower bound algorithmically?
Such an algorithm would allow one to use lower depth quantum circuits to still achieve a speedup over classical methods (albeit with a slightly decreased speedup, due to the increased parallel runs required), allowing one to push quantum speedups towards the NISQ era.

A second line of recent works have demonstrated this is possible, using differing techniques to reduce the circuit depth.
First, Burchard \cite{burchard2019lower} suggested an algorithm which nearly matches his lower bound in certain settings.
In particular, if it is easy to split one's Monte Carlo problem into $p$ equally sized subproblems, then one can achieve a $\tilde{\O}(\sqrt{p})$-fold reduction in the depth by running the quantum Monte Carlo algorithm separately on each subproblem and classically combining the results\footnote{More formally, one runs the quantum Monte Carlo algorithm separately on each subproblem to $ \O(\sqrt{p}/\epsilon)$ relative accuracy (which, as the domain is $p$-fold smaller, results in $ \O(1/\epsilon \sqrt{p})$ absolute accuracy in the valuation), and then classically averaging results. The random walk behavior of averaging these errors ensures a total accuracy of $ \O(1/\epsilon)$ as desired. }. 
This asymptotically matches the lower bound up to logarithmic factors, realizing the smooth continuum between quantum and classical Monte Carlo in the asymptotic setting.
However, this method has several limitations which might make it challenging to implement in practice.
First, the method incurs an additional $ \O(\log p)$ slowdown relative to the lower bound, due to the need to amplify the probability that each of the $p$ counts lands in its respective confidence interval.
In the setting of low-depth devices, $p$ will need to be rather large to allow for low-depth circuits, so this term could become particularly significant.
Second, this method assumes the Monte Carlo problem can be easily split into subproblems --- which might not be the case in certain implementations of the quantum algorithm\footnote{In more detail, if the quantum sample is obtained by ``quantizing'' a \emph{digital} classical sampling algorithm --- i.e. by creating a uniform superposition of random seeds $\frac{1}{\sqrt{|R|}}\sum_{r\in R} \ket{r}$, and computing the classical sampling algorithm in superposition over those seeds --- then the subdivision can be obtained by dividing the random seeds into $p$ random buckets. But if the quantum sample is obtained from purely \emph{analog} methods --- i.e. by directly creating the superposition over market histories using quantum gates --- then it may be unclear how to split the problem into $p$ disjoint pieces. Note that one could not simply postselect on the market histories falling into random buckets of market outcomes, since in general reflecting about such a conditional sample state would incur an additional $ \O(\sqrt{p})$-fold cost in the algorithm by oblivious amplitude amplification \cite{berry2014exponential}, thus eliminating the reduction in depth achieved.}, particularly if the quantum sample is prepared using analog methods like quantum data loaders \cite{patent2} or Grover-Rudolph \cite{grover2002creating,rebentrost2018quantum}.

Second, Kerenidis and Prakash \cite{KPpatent} have developed a different approach to smoothly interpolate between quantum and classical Monte Carlo algorithms without these limitations, opening the door to quantum speedups at even lower circuit depths.
In particular, Kerenidis and Prakash showed that one can use many runs of low-depth quantum circuits to accelerate Monte Carlo evaluation, where the speedup achieved is proportional to the depth of the quantum circuits evaluated, but without any need to split into subproblems.
The algorithm is applicable for amplitude estimation in general and does not require additional structure in the quantum sampler --- opening the door to NISQ-oriented quantum sample preparation methods such as quantum data loaders --- while simultaneously achieving the optimal tradeoff. 

In particular, Kerenidis and Prakash's algorithm builds on the framework of Suzuki et al. \cite{suzuki2020amplitude} for QFT-free amplitude estimation.
In Suzuki et al.'s approach, the basic idea is to run many runs of the quantum Monte Carlo algorithm on the entire domain (and without the QFT) with varying depths, and then perform a classical maximum likelihood postprocessing algorithm to estimate the desired mean $V$ after the fact. 
In particular they run the algorithm at depths which follow an exponential schedule, i.e. at depths $1,2,4,8\ldots$ all the way up to the maximum depth of $ \O(1/\epsilon)$.
This approach does not provide a depth reduction, but does eliminate the need for a QFT at the end of the algorithm\footnote{Though we note it remains unclear if Suzuki et al.'s max-likelihood postprocessing will remain efficient asymptotically \cite{aaronson2020quantum}.}.
Kerenidis and Prakash, on the other the other hand, consider running a similar algorithm where the depths follow a polynomial (and indeed even sublinear) schedule, with a maximum depth of $D\ll  \O(1/\epsilon)$.
This provides a significant depth reduction to the algorithm, at a cost of more parallel runs.
They showed that by using carefully chosen sub-linear schedules for depth reduction, where the depth schedules can also be selected \emph{adaptively}, one can achieve the optimal speedup for the given depth, namely it tightly matches Burchard's bound.
Interestingly Burchard's bound does not immediately apply to the Kerenidis-Prakash algorithm, and we leave it as an open problem to determine if it can be generalized to this case\footnote{This is because Burchard's bound is for non-adaptive parallel quantum sample algorithms, where each run of the algorithm has the same depth, rather than (classically) adaptive runs of varying depths.}.
Subsequently Tanaka et al. \cite{tanaka2020amplitude} have explored similar algorithmic ideas, in particular analyzing super-linear depth schedules in the presence of depolarizing noise providing evidence that such methods can be robust against noise.
It is interesting to explore to what extent a quantum speedup can be realized with other depth schedules and noise models as well.

\begin{figure}
    \centering
\begin{tikzpicture}
\node[align=center] at (0,-0.25) {$1$};
\node[align=center] at (3,-0.25) {$\frac{1}{\epsilon^{1/2}}$  };
\node[align=center] at (6,-0.25) {$\frac{1}{\epsilon}$};
\node[align=center] at (9,-0.25) {$>\frac{1}{\epsilon}$ };
\node[align=center] at (6,3.75) {Max speedup\\ achieved};
\node[align=center] at (6,-1.25) {Logical depth of quantum circuit\\(number of quantum samples in series)};
\node[align=center] at (-1.5,2.5) {Standard \\quantum \\algorithm};
\node[align=center] at (-1.5,0.75) {K-P framework};
\fill[blue] (6,2) rectangle (12,3);
\draw (0,2) rectangle (12,3);
\shade[left color=white,right color=blue] (0,0.25) rectangle (6,1.25);
\fill[blue] (6,0.25) rectangle (12,1.25);
\draw (0,0.25) rectangle (12,1.25);
\draw[dashed] (6,2) -- (6,3);
\draw[dashed] (6,0.25) -- (6,1.25);
\end{tikzpicture}    
    
    \caption{Schematic of the Kerenidis-Prakash framework. Suppose one wishes to perform a Monte Carlo calculation to some fixed accuracy $\epsilon$. In the standard quantum algorithm for Monte Carlo, no speedup is achieved until one can perform circuits of a fixed depth --- namely the depth of $1/\epsilon$ Grover iterates in series. From that circuit depth onwards (dashed line), one obtains a speedup factor of $1/\epsilon$ over classical computing. In contrast the Kerenidis-Prakash framework smoothly interpolates between the classical and quantum algorithms, yielding a speedup proportional to the depth of the quantum circuits that can be run (and matches the speedup at depth $1/\epsilon$). }
    \label{fig:MCresourceschematic}
\end{figure}
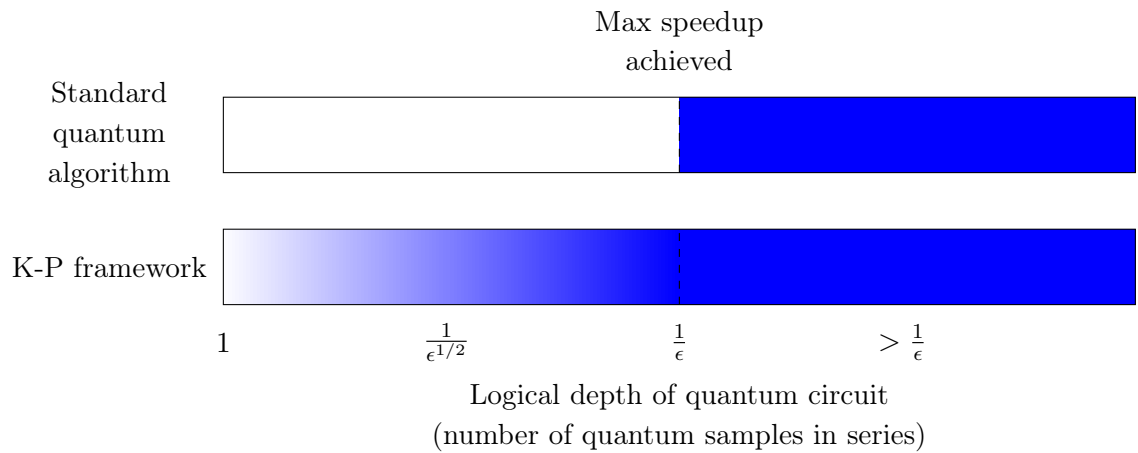

These low-depth speedups are represented schematically in Figure \ref{fig:MCresourceschematic}, where the speedup achieved is plotted vs the number of serial ``quantum samples'' from $S,f$ required to run both the standard quantum algorithm and the Kerenidis-Prakash variant. We see that the standard quantum algorithm starts offering a speedup of $ \O(1/\epsilon)$ once the depth becomes $ \O(1/\epsilon)$, but does not generate speedups at lower depths. 
Therefore in some sense the standard quantum algorithm is ``all or nothing'' when one wishes to price a derivative to a fixed value of $\epsilon$ --- one can either perform a circuit of depth $1/\epsilon$ and get the full speedup, or else one cannot run the algorithm at all\footnote{Of course one can always run the algorithm at a lower value of $\epsilon' \gg \epsilon$ (with lower depth). But is unclear how this will help if one wishes to price the derivative to accuracy $\epsilon$}. 
What the Kerenidis-Prakash algorithms do (as well as the Burchard algorithm in certain settings) is substantially extend the regime of quantum speedup starting now from depth $ \O(1)$ and all the way to depth $ \O(1/\epsilon)$, where now the speedup grows linearly with the depth and matches the standard algorithm when the depth reaches $ \O(1/\epsilon)$.
This saturates the Burchard bound, showing that it is tight not only asymptotically in particular settings \cite{burchard2019lower}, but also in more general amplitude estimation settings, and with improved pre-factors.

To be more precise, the Kerenidis-Prakash algorithm is parameterized by a number $\beta$ which takes values between 0 and 1 and interpolates between the classical algorithm (when $\beta=1$, we only perform a single application of $S$ and $f$ and need to repeat the algorithm $ \O(1/\epsilon^2)$ times) and the standard quantum algorithm (when $\beta=0$, we perform $ \O(1/\epsilon)$ applications of $S$ and $f$ in series and we only repeat the algorithm once). Taking values of $\beta$ between 0 and 1 provides lower depth circuits at the expense of more oracle calls while still retaining an asymptotic advantage over classical Monte Carlo methods. Let us assume that a single run of the quantum circuit consisting of the stochastic market model $S$ followed by the payoff function $f$ (often referred to as a call to the {\em quantum oracle}) uses $n$ qubits and has circuit depth $d$. In Table \ref{table1}, we summarize the necessary resources for the different quantum algorithms, in particular the total number of qubits, the maximum circuit depth one can run without expecting errors, and the total number of calls to the quantum circuit consisting of the stochastic market model $S$ followed by the payoff function $f$.

\begin{table} 
\begin{center} 
\small{
  \begin{tabular}{|l|c|c|c|} 
   \hline
  &&&\\
   \textbf{Algorithm} & \textbf{Qubits} & \textbf{Depth} & \textbf{Number of calls}  \\
      &&&\\
    \hline
      &&&\\
      Amplitude Estimation \cite{brassard1998quantum}& $n + \log(1/\epsilon)$ & $d \cdot \frac{1}{\epsilon}+\log\log(1/\epsilon)$ & $\frac{1}{\epsilon}$ \\
    &&&\\
    \hline
     &&&\\
    QFT-free Amplitude Est. \cite{suzuki2020amplitude, aaronson2020quantum}  & $n$ & $d \cdot \frac{1}{\epsilon}$ & $\frac{1}{\epsilon}$ \\
      &&&\\
    \hline 
     &&&\\
     Parallel Approximate Counting \cite{burchard2019lower}    & $n$ &  $d \cdot \left(\frac{1}{\epsilon}\right)^{1-\beta} \log(1/\epsilon)$ & $ \left( \frac{1}{\epsilon} \right)^{1+\beta}\log(1/\epsilon)$ \\
    &&&\\
    \hline 
     &&&\\
      Kerenidis-Prakash method \cite{KPpatent}  & $n$ &  $d \cdot \left(\frac{1}{\epsilon}\right)^{1-\beta}$ & $ \left( \frac{1}{\epsilon} \right)^{1+\beta}$  \\
    &&&\\
    \hline 
   \end{tabular}
   \caption {Tradeoffs for different amplitude estimation algorithms.} \label{table1} 
   }
  \end{center} 
   \end{table}

\begin{figure}
    \centering
   \begin{tabular}{ |c | p{1cm}| p{1.1cm}|p{1.3cm}|p{1.1cm}|p{1.1cm}|p{1.3cm}| p{1.1cm} | p{1.1cm} | p{1.3cm} | p{1.1cm} |c| }
 \hline
 \multicolumn{1}{|c|}{} & \multicolumn{10}{|c|}{Kerenidis-Prakash framework} \\
 \hline
  \multicolumn{1}{|c|}{} & \multicolumn{1}{p{1.1cm}}{classical $\beta=1$} 
  & \multicolumn{3}{|c|}{$\beta=2/3$} & \multicolumn{3}{|c|}{$\beta=1/3$} & \multicolumn{3}{p{2.6cm}|}{ \small{standard quantum} $\beta=0$} \\
 \hline
 
 $\epsilon$ & \small{classical samples} & \small{q. samples in series} & allowed sample error rate & \small{speedup} & \small{q. samples in series}  & allowed sample error rate & \small{speedup} & \small{q. samples in series} & allowed sample error rate &\small{speedup} \\
 \hline
 $10^{-2}$   & $10^4$ &   $ 11$ & $9\times 10^{-4}$ & $ 5$ & $ 45$ & $2\times 10^{-4}$&$22$ & $2\times10^2$ & $5\times10^{-5}$ & $10^2$\\

 $10^{-3}$ &  $10^6$    & $21$ & $5\times 10^{-4}$ &  $10$ & $2 \times 10^2$ & $5\times10^{-5}$ & $ 1 \times 10^2$ & $2\times10^3$& $5\times10^{-6}$& $10^3$\\
 
 $10^{-4}$ &  $ 10^8$      &  $45$ & $2\times10^{-4}$ & $ 22$ & $9\times10^2$ & $1\times10^{-5}$ & $5\times 10^2$ & $2\times10^4$& $5\times10^{-7}$& $10^4$\\
 
 $10^{-5}$ &$10^{10}$   &  $93$ & $1\times 10^{-5}$ & $46$ & $ 4 \times 10^3$& $2\times 10^{-6}$ &$ 2 \times 10^3$ & $2\times10^5$& $5\times 10^{-8}$& $10^5$\\

 \hline
\end{tabular}
    \caption{Back-of-the-envelope resource estimates for pricing an option using quantum algorithms for Monte Carlo --- both the standard algorithm and Kerenidis-Prakash's variants for $\beta=\{3/4,1/2,1/4\}$. Here we are counting the number of classical and quantum samples needed to apply each algorithm.
    We compute the raw quantum speedup obtained --- i.e. the ratio of the number of classical samples needed to the number of quantum samples needed by each algorithm --- as well as the error rate permissible in the quantum sampling algorithm to achieve a $99\%$ confidence interval at the end of the algorithm. In these calculations, we take an asymptotic viewpoint, the actual speedups for the quantum algorithm may be better or worse by a constant factor. }
    \label{fig:MCresource1}
\end{figure}

To see the speedup vs. depth tradeoffs of this algorithm more concretely, in Figure \ref{fig:MCresource1} we calculate the quantum speedup achieved for different accuracy parameters $\epsilon$ and various settings of the interpolation parameter $\beta$. 
Here we provide the number of classical samples needed, as well as the speedup offered and the necessary depth (counted as the number of quantum samples in series) of different quantum algorithms.
We emphasize that these are back of the envelope calculations only, but we believe they clearly illustrate certain trends.
Namely, we can clearly see that the Kerenidis-Prakash framework of quantum Monte Carlo algorithms offers speedups from modest depths and in general optimally trades a reduction in the depth of the quantum circuits for a reduction in the speedup obtained by the algorithm. 
Note that here we are putting aside the depth or the time that one needs to perform a single application of the stochastic market model $S$ followed by the payoff function $f$. In other words, we are assuming that taking a ``sample'' has unit cost for now for both the quantum and classical case. This provides us with an optimistic bound on the quantum speedup obtainable by the algorithm, as in reality the time/resources required to perform a quantum application of $S$ and $f$ may be higher than that required for a classical application --- a fact we will explore in more detail in the next section when we estimate what is needed to make these speedups practical.

\subsection{Resource estimates for quantum Monte Carlo methods}
\label{sec:MCresourceestimates}

In this section we ask the following question:  when will quantum Monte Carlo methods become practically relevant? And to what extent will these new, lower-depth quantum MC algorithms bring this speedup closer to reality? While we will not address the issue of ``how many years'' due to the high uncertainties in future hardware progress, in this section we will provide a coarse estimate of the quantum resources required to run the standard algorithm and the Kerenidis-Prakash variants to achieve a practical speedup over a classical CPU.
In particular we will estimate the effective error rate and clock speed required to achieve a speedup using these algorithms.
We find that the Kerenidis Prakash variant opens up a new sector of the hardware landscape at which quantum speedups is possible, which is ``NISQ-ier'' (i.e. closer to near term quantum hardware) than the standard quantum Monte Carlo algorithm.
Nevertheless, these hardware parameters are still well beyond the capabilities of current prototype quantum computers.

The two key parameters which control the extent of quantum speedup achievable with quantum Monte Carlo algorithms are the effective clock speed and the effective error rate of the quantum computer.

The effective clock speed of the quantum computer determines how long it takes to perform a ``quantum sample'' with the device.
In our speedup calculations in the previous section, we optimistically assumed that the time taken to perform a quantum sample from $S,f$ is the same as the time required to take a classical sample.
However, in reality the clock speed of current quantum technologies is slower than that of classical CPUs.
For example, current superconducting qubit systems can perform 2-qubit gates in roughly $20$ns, i.e. their clock speed is roughly 20 times slower than classical CPUs operating at $\sim1$GHz.
(see \cite{campbell2019applying} for a discussion).
This slowdown could be more significant if the algorithm requires quantum error correction to implement, as in error corrected systems the effective clock speed is slower than the physical clock speed of the device due to overheads in error correction (see e.g. \cite{gidney2019factor}).

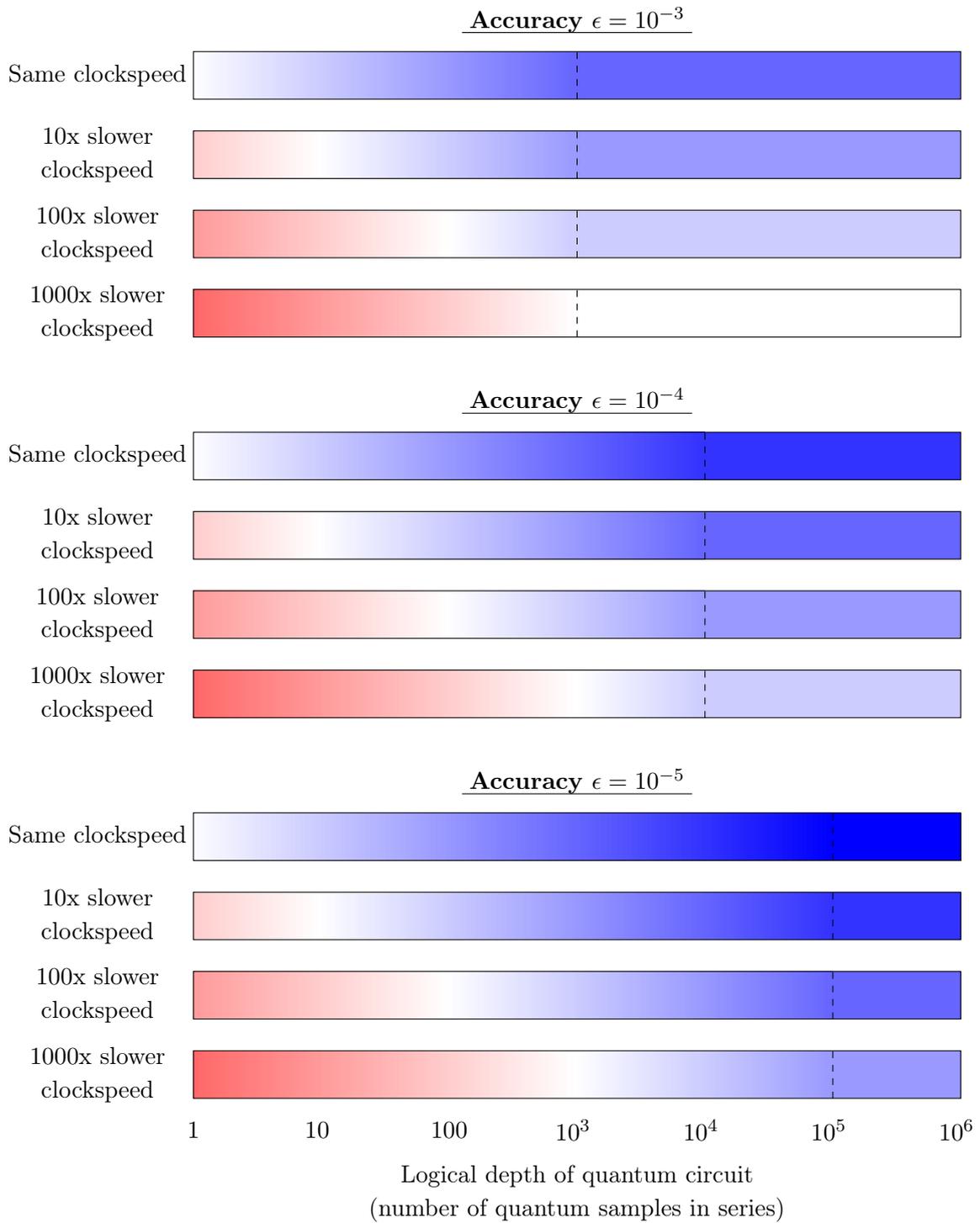
\begin{figure}
    \centering
\begin{tikzpicture}

\node[align=center] at (6,4.0) { \textbf{Accuracy} $\epsilon=10^{-3}$};
\draw (4.2,3.8) -- (7.8,3.8);

\shade[left color=white,right color=blue!60!white] (0,2.75) rectangle (6,3.5);
\fill[blue!60!white] (6,2.75) rectangle (12,3.5);
\node[align=center] at (-1.5,3.125) {Same clockspeed};
\draw[dashed] (6,2.75) -- (6,3.5);
\draw (0,2.75) rectangle (12,3.5);

\shade[left color=red!20!white, right color=white] (0,1.5) rectangle (2,2.25);
\shade[left color=white,right color=white!60!blue] (2,1.5) rectangle (6,2.25);
\fill[white!60!blue] (6,1.5) rectangle (12,2.25);
\node[align=center] at (-1.5,1.875) {$10$x slower \\clockspeed};
\draw[dashed] (6,1.5) -- (6,2.25);
\draw (0,1.5) rectangle (12,2.25);

\shade[left color=red!40!white, right color=white] (0,0.25) rectangle (4,1.0);
\shade[left color=white,right color=white!80!blue] (4,0.25) rectangle (6,1.0);
\fill[white!80!blue] (6,0.25) rectangle (12,1.0);
\node[align=center] at (-1.5,0.625) {$100$x slower \\clockspeed};
\draw[dashed] (6,0.25) -- (6,1.0);
\draw (0,0.25) rectangle (12,1.0);

\shade[left color=red!60!white, right color=white] (0,-1) rectangle (6,-0.25);
\draw[dashed] (6,-1) -- (6,-0.25);
\node[align=center] at (-1.5,-0.625) {$1000$x slower \\clockspeed};
\draw (0,-1) rectangle (12,-0.25);

\begin{scope}[shift={(0,-6)}]

\node[align=center] at (6,4.0) { \textbf{Accuracy} $\epsilon=10^{-4}$};
\draw (4.2,3.8) -- (7.8,3.8);

\shade[left color=white,right color=blue!80!white] (0,2.75) rectangle (8,3.5);
\fill[blue!80!white] (8,2.75) rectangle (12,3.5);
\node[align=center] at (-1.5,3.125) {Same clockspeed};
\draw[dashed] (8,2.75) -- (8,3.5);
\draw (0,2.75) rectangle (12,3.5);

\shade[left color=red!20!white, right color=white] (0,1.5) rectangle (2,2.25);
\shade[left color=white,right color=white!40!blue] (2,1.5) rectangle (8,2.25);
\fill[white!40!blue] (8,1.5) rectangle (12,2.25);
\node[align=center] at (-1.5,1.875) {$10$x slower \\clockspeed};
\draw[dashed] (8,1.5) -- (8,2.25);
\draw (0,1.5) rectangle (12,2.25);

\shade[left color=red!40!white, right color=white] (0,0.25) rectangle (4,1.0);
\shade[left color=white,right color=white!60!blue] (4,0.25) rectangle (8,1.0);
\fill[white!60!blue] (8,0.25) rectangle (12,1.0);
\node[align=center] at (-1.5,0.625) {$100$x slower \\clockspeed};
\draw[dashed] (8,0.25) -- (8,1.0);
\draw (0,0.25) rectangle (12,1.0);

\shade[left color=red!60!white, right color=white] (0,-1) rectangle (6,-0.25);
\shade[left color=white, right color = white!80!blue] (6,-1) rectangle (8,-0.25);
\fill[white!80!blue] (8,-1) rectangle (12,-0.25);
\draw[dashed] (8,-1) -- (8,-0.25);
\node[align=center] at (-1.5,-0.625) {$1000$x slower \\clockspeed};
\draw (0,-1) rectangle (12,-0.25);
\end{scope}

\begin{scope}[shift={(0,-12)}]
\node[align=center] at (6,4.0) { \textbf{Accuracy} $\epsilon=10^{-5}$};
\draw (4.2,3.8) -- (7.8,3.8);

\shade[left color=white,right color=blue] (0,2.75) rectangle (10,3.5);
\fill[blue] (10,2.75) rectangle (12,3.5);
\node[align=center] at (-1.5,3.125) {Same clockspeed};
\draw[dashed] (10,2.75) -- (10,3.5);
\draw (0,2.75) rectangle (12,3.5);

\shade[left color=red!20!white, right color=white] (0,1.5) rectangle (2,2.25);
\shade[left color=white,right color=white!20!blue] (2,1.5) rectangle (10,2.25);
\fill[white!20!blue] (10,1.5) rectangle (12,2.25);
\node[align=center] at (-1.5,1.875) {$10$x slower \\clockspeed};
\draw[dashed] (10,1.5) -- (10,2.25);
\draw (0,1.5) rectangle (12,2.25);

\shade[left color=red!40!white, right color=white] (0,0.25) rectangle (4,1.0);
\shade[left color=white,right color=white!40!blue] (4,0.25) rectangle (10,1.0);
\fill[white!40!blue] (10,0.25) rectangle (12,1.0);
\node[align=center] at (-1.5,0.625) {$100$x slower \\clockspeed};
\draw[dashed] (10,0.25) -- (10,1.0);
\draw (0,0.25) rectangle (12,1.0);

\shade[left color=red!60!white, right color=white] (0,-1) rectangle (6,-0.25);
\shade[left color=white, right color = white!60!blue] (6,-1) rectangle (10,-0.25);
\fill[white!60!blue] (10,-1) rectangle (12,-0.25);
\draw[dashed] (10,-1) -- (10,-0.25);
\node[align=center] at (-1.5,-0.625) {$1000$x slower \\clockspeed};
\draw (0,-1) rectangle (12,-0.25);

\node[align=center] at (0,-1.5) {$1$};
\node[align=center] at (2,-1.5) {$10$ };
\node[align=center] at (4,-1.5) {$100$};
\node[align=center] at (6,-1.5) {$10^3$ };
\node[align=center] at (8,-1.5) {$10^4$ };
\node[align=center] at (10,-1.5) {$10^5$ };
\node[align=center] at (12,-1.5) {$10^6$ };

\node[align=center] at (6,-2.5) {Logical depth of quantum circuit\\(number of quantum samples in series)};

\end{scope}
\end{tikzpicture}    
    \caption{Schematic of how the K-P framework quantum speedup behaves when one takes into account slower clock speeds, for different values of $\epsilon$. Red indicates slowdowns, white is neutral, and blue indicates speedups, and the strength of the color indicates the degree of speedup/slowdown. The depth at which maximum speedup is achieved is denoted by a dashed line. }
    \label{fig:MCslowdowns}
\end{figure}

Therefore one should expect that a future quantum device will operate at an initial ``slowdown'' relative to classical devices.
This must be compensated for by the corresponding quantum algorithmic speedup obtained by the algorithm in order to obtain a practical speedup.
To see this diagrammatically, in Figure \ref{fig:MCslowdowns} we plot the practical speedup obtained by the Kerenidis-Prakash framework for a variety of accuracy parameters $\epsilon$ as a function of the effective slowdown in the quantum clock speed (10x, 100x, and 1000x). While a slower quantum clock erodes part of the quantum speedup obtained, one can clearly see certain parameter regimes of commercial relevance (and more so for larger accuracies).

The second parameter that controls the ability to obtain a practical speedup is the effective error rate of the quantum device.
As discussed, the high noise rates present in current quantum computers limit the depth of the quantum computations that can be performed without errors. 
Therefore quantum machines with high error rates will be unable to perform deep variants of the Kerenidis-Prakash algorithm, as they will be unable to perform many ``quantum samples'' in series.
Since the speedup obtained is proportional to the number of quantum samples performed in series, this means that higher-error rate machines will be more limited in the speedups they can obtain using these algorithms.

One can therefore explore the interplay between these parameters --- lower error rate machines can provide larger quantum speedups (as they can run deeper circuits), and faster clock speed quantum devices can also more easily generate quantum speedups (as they suffer less initial ``slowdown'' from clock speed). Therefore quantum speedups are obtainable in certain hardware parameter regimes but not in others. Ideally, one will have in the future quantum machines with both low error rates and high clock speed. Nevertheless, speedups may still be obtainable in slow clock speed regimes, if, at the same time, the corresponding effective error rate is considerably low. Broadly speaking, this is the regime in which future, large-scale, error-corrected quantum devices are expected to operate, due to the slowdowns in clock speedups required for quantum error correction.
On the other hand, speedups may also be obtainable by quantum machines with higher error rates, so long as their clock speeds are faster.
Broadly speaking, near-term prototype quantum computers (i.e. NISQ devices) fall into this parameter regime, since they do not have slowdowns due to error correction, but as a consequence they suffer from higher error rates.

\begin{figure}
    \centering
    \includegraphics[scale=0.75]{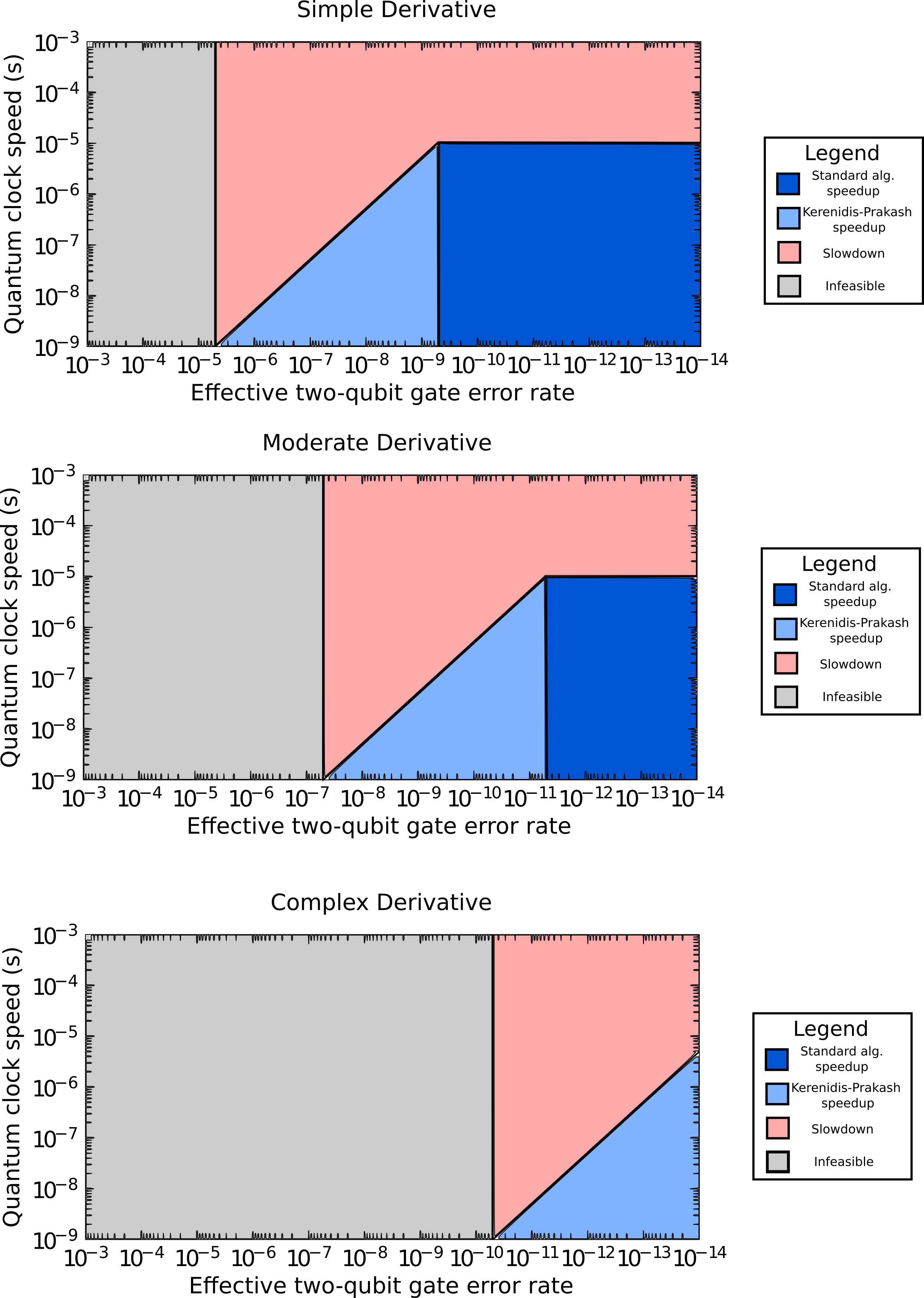}

    \caption{Estimates of the quantum hardware parameters required to run the K-P algorithm and still achieve a practical speedup for accuracy $\epsilon=10^{-4}$. We display the effective error rate needed to run the algorithm, as well as the effective (possibly error-corrected) clockspeed required to obtain a quantum speedup. Red areas indicate slowdowns, gray areas are infeasible to run the algorithm due to too high of error rates, light blue indicates a speedup under the K-P framework only, and blue indicates a quantum speedup under the standard quantum algorithm. 
    Simple derivatives corresponds to quantum samplers with $10^3$ gates total, moderate derivatives to $10^5$ gates total, and complex derivatives to $10^8$ gates total. We assume a 1GHz quantum processor performs a quantum sample in the same time a classical CPU can produce a classical sample. }
    
    \label{fig:MChardwareparameterspace}
\end{figure}

In Figure \ref{fig:MChardwareparameterspace} we translate this tradeoff into concrete hardware parameters.
In particular we delineate which combinations of effective quantum error rate and quantum clock speed could be used to obtain practical speedups for Monte Carlo algorithms.
Since quantum error rates are commonly quoted in terms of an error rate per two qubit gate, these tradeoffs depend in part on the complexity of the derivative being priced or the VaR calculation being performed (since more complex derivatives/VaR calculations require more two-qubit gates to perform a single quantum sample).
We therefore show this tradeoff for a variety of complexities of derivatives to be priced --- ranging from very simple derivatives like a call option on lognormal random walks to highly complex instruments requiring tens of millions of logical operations to implement the stochastic market model and payoff function. 
We estimate that the simplest derivatives --- like a European call option on a single security on a lognormal random walk --- might require merely $10^3$ Toffoli gates to perform a quantum sample, though we note that this type of derivatives can easily be priced classically by the Black-Scholes formula.
On the other hand, more complicated securities may require significantly more complicated quantum sample subroutines.
Forthcoming work of Chakrabarti et al. \cite{goldmaninprep} suggests that classically non trivial derivatives may require $\sim 10^5$ Toffoli gates per quantum sample, and more complex derivatives may require up to $10^8$ Toffoli gates per quantum sample. One may hope to improve these numbers by finding even better quantum ways of preparing the samples.
We therefore quote the effective error rates needed to run this algorithm for simple ($10^3$ gates per quantum sample), moderate ($10^5$ gates per quantum sample) and complex ($10^8$ gates per quantum sample) derivatives.

Please note that we are quoting these error rates in terms of the \emph{effective} error rates and clock speeds of the quantum device --- i.e. the error rate and clock speed may refer to real quantum hardware operating with physical qubits or obtained after error correction has been applied (if necessary). With error correction the effective clock speed could be slower than the physical clock speed by orders of magnitude.
We emphasize that while our estimates are ``back of the envelope'' rather than very precise attempts to quantify the actual costs in hardware, we believe they are still informative estimates of the resources required.

This figure reveals clear trends.
First, above a certain error rate, the algorithm is infeasible to run (gray area), simply because one is incapable of running the stochastic market model $S$ and payoff function $f$ in superposition with high fidelity\footnote{Here we fix 99\% as the target fidelity of the quantum algorithm.}.
Therefore with these algorithms there is a certain ``ante'' in terms of quantum error rate that one must be able to achieve before the algorithm is even applicable to the problem at hand.
This ``ante'' is particularly high for complex derivatives, but lower for simpler derivatives which admit NISQ state preparation procedures.
Second, to run the standard quantum Monte Carlo algorithm and get the maximum speedup, one must have a very small error rate, as one must perform many quantum samples in series. Within this region with small effective error rates (dark blue region), the algorithm can still obtain speedups with slower effective clock speeds.
Third, one can see that the Kerenidis-Prakash framework ``unlocks'' a new region of the parameter space for quantum speedups --- in particular a region which allows for higher effective error rates, but demands faster quantum clock speeds.
This parameter regime is more ``NISQ-friendly'' than the standard quantum algorithm, but is still currently beyond the capabilities of the available quantum computers.
Finally, one should note that there is a ``dead zone'' in the chart for which no quantum speedup is possible with this algorithm (red region)-- in particular hardware with very slow effective quantum clock speeds, or very high error rates, precludes the possibility of a quantum speedup.

The approach of Kerenidis and Prakash, while it sacrifices some of the quantum speedup, does successfully decrease the stringent error correction requirements of the standard algorithm.
In particular, it could place the first practical speedups for Monte Carlo --- in a specialized parameter regime of relatively simple payoff functions and relatively high desired accuracy $\epsilon$ --- within reach of ``intermediate-scale'' devices capable of error rates of merely $10^{-5}-10^{-7}$ (compared to current physical error rates of $10^{-3}$). Such error rates are not available now, but they might be possible in the intermediate term, and certainly they are substantially easier to obtain in hardware than say the error rates of $\sim10^{-9}$ required to run Shor's algorithm \cite{gidney2019factor} or the original quantum algorithm for Monte Carlo \cite{montanaro2015quantum}.
On the other hand, running this algorithm to price highly complex securities may require \emph{even better} gate accuracies than are required to run Shor's algorithm to break modern internet security, which is estimated to be decades away.
Therefore, we expect that NISQ quantum computers are more likely to provide advantages for pricing simpler derivatives with high desired accuracy than 
pricing highly complex derivatives which remains a longer term prospect.

The timeline for the future availability of quantum hardware with these performance specifications is a major open question.
Present day hardware with physical error rates of $10^{-2}$-$10^{-3}$ and clock speeds at least 20 times slower than classical CPUs lie on the left vertical axis in Figure \ref{fig:MChardwareparameterspace} --- i.e. it is not feasible to run these quantum algorithms right now on current hardware in any commercially meaningful way, even for simple derivatives.
An improvement in the \emph{quality} of the physical qubits (i.e. the base error rate)
would shift this base point towards the blue area that represents the regime where quantum speedups are expected. 
On the other hand, an improvement in the \emph{quantity} of physical qubits (at a fixed error rate) would allow for the implementation of quantum error correction. 
This error correction will push the hardware parameters to the right (better error rates), but also upwards as error correction decreases the effective clock speed.  
If we draw a line extrapolating the development of error corrected hardware, it can be seen from the figure that the slope of the line will be critical in determining the extent of quantum Monte Carlo speedups achievable by them. 
A combination of improvements in qubit quality and improved error correction may therefore be needed to achieve significant quantum speedups for Monte Carlo methods.

Since the trajectory of error correction is also dependent on the base error rate, improving qubit \emph{quality} may be especially important for this algorithm.
For example, under more conservative projections of the physical noise rate (say the noise rate is $10^{-3}$ per two qubit gate), any variant of this algorithm --- even for pricing simple securities --- would require quantum error correction to implement, substantially slowing the effective clock speed and potentially requiring orders of magnitude more physical qubits to implement.
Indeed Egger et al.\ \cite{egger2019credit} recently used such hardware projections to estimate that millions of physical qubits would be required to run the standard quantum algorithm at scale at such base error rates.
The fact that the effective clock speeds of such devices would be significantly slowed \cite{gidney2019factor} could decrease the utility of this algorithm.
On the other hand, the prospects for a practical speedup improve significantly under more optimistic projections of the physical error rate of quantum devices.
For instance, it has been suggested that topological quantum computers could be capable of extremely low error rates \cite{reiher2017elucidating} (see also \cite{campbell2019applying} for a similar discussion).
Under such scenarios this algorithm may only require limited or no error correction to run, particularly for pricing relatively simple derivatives.
Moreover, when error correction is needed, the clockspeed/error rate tradeoff improves significantly if one has lower underlying physical error rates.
Hence qubit \emph{quality} may be a controlling factor in realizing practical speedups from this algorithm.\\

In summary, the low-depth quantum Monte Carlo methods expand the possibility of quantum speedups to much larger parameter regimes, that one can start approaching as faster, better-quality gates become available (either from lower physical error rates or from the implementation of error correction). Given the uncertainty in the timeline for the development of quantum hardware the answer to the question of when we will be able to run quantum Monte Carlo methods at scale will continue to evolve both with the advent of new hardware and with better algorithms and error correcting codes. 
Despite the fact that current hardware cannot match the requirements of even the low-depth Monte Carlo methods we described, the design of novel Monte Carlo algorithms tailored to the hardware specifications is necessary in order to keep bringing forward the date at which the quantum algorithms become realizable and provide impact on the finance industry.

\section{Portfolio Optimization}
\label{sec:portfolioopt}

A second application of quantum algorithms in finance is to portfolio optimization, a classic problem in mathematical finance going back to Markowitz \cite{markowitz1952}. 
Suppose one is allocating investment amongst $n$ assets of differing returns, and whose returns are correlated in some known way.
The goal is to minimize the risk of the portfolio while achieving a target rate of return.
In its simplest form the problem is relatively easy to solve, as it reduces to solving a single linear system of equations.
However the computational difficulty of this problem increases when one adds more realistic constraints to the problem.
For instance, one might wish to enforce positivity constraints (so that one can only buy and not sell assets), or integer constraints (so that one can only invest in assets in fixed increments), or upper limits on the amount one might invest in a particular asset (due e.g.\ to limited liquidity).
Additionally the returns or correlations between the assets may vary in time, but there may be transaction costs to buying/selling assets, and therefore the optimal trading trajectory may not be locally optimal at each time. 
With these more realistic constraints, the portfolio optimization problem can become substantially more challenging to solve, even approximately, and is a staple of modern computational finance.

A variety of quantum algorithmic approaches have been developed to solve, or approximately solve, several variants of the portfolio optimization section, with varying degrees of quantum speedups.
In this section we briefly review these quantum algorithmic approaches, and also describe recent works of QC Ware and collaborators to develop and improve upon them. 
As in the previous section, our focus will be on developing methods which bring quantum speedups closer to fruition with lesser hardware requirements.

\subsection{Three variations of the portfolio optimization problem}

Let us first review the mathematical definition of the portfolio optimization problem, and how different variants of the problem have different degrees of computational difficulty.
We consider $n$ assets and we define $R(t) \in \R^n$ to be the (column) vector of returns for all assets at time $t$. 
The expected return $\mu \in \R^n$ and the covariance matrix $\Sigma\in \R^{n\times n}$ of asset returns are as follows: 
\begin{align*} 
\mu &= \frac{1}{T} \sum_{t \in [T]} R(t) \\
\Sigma &= \frac{1}{T-1}  \sum_{t \in [T]} ( R (t) - \mu )   ( R (t) - \mu )^\intercal
\end{align*}
A portfolio $w$ is a weight vector $w=\left(w_{1}, \dots, w_{n}\right)^\intercal \in \R^n$, where $w_j$ specifies the investment on asset $j$, and the return $R_w(t)$ of the portfolio $w$ at time $t$ is given by: 
\begin{align*}
    R_w(t) & =\sum_{i=1}^{n} w_{i} R_{i}(t)=w^\intercal R(t).
\end{align*}
The expected return $\mu_w \in \R$ of the portfolio $w$ is equal to 
\begin{align*}
\mu_w & =\E[R_w]=w^\intercal \mu,
\end{align*}
and its variance $\sigma_w^2\in\R$ is given by
\begin{align*}
    \sigma_w^{2}& =\E\left[(R_w-\mu_w)^{2}\right]=w^\intercal \Sigma w
\end{align*}
For Markowitz, the objective of an investor is to achieve the lowest level of risk for a given return objective. There are ways to choose such a return objective for the assets, and Black and Litterman in \cite{black1992litterman} provide a variety of such methods, also taking into account the investor views about the future performance of the portfolio assets. In high level, the investor's objective can therefore be expressed as a Mean-Variance optimization problem. 

The standard Markowitz Mean-Variance approach resolves the unconstrained portfolio optimization problem, which, given $\Sigma, R, A, \mu, b$, is defined as follows:
\begin{align*}
\min _{w\in\R^n} & w^\intercal \Sigma w \\ \text { subject to } & R^\intercal w=\mu \\ &A^\intercal w = b \end{align*}
The first constraint corresponds to fixing the return, while one may add more equality constraints of the form $A^\intercal w=b$, for example the budget constraint $\mathbf{1}^\intercal w=1$ that implies that the wealth is entirely invested. 

Of course, the above description of the portfolio optimization problem is not unconstrained, since it has, in fact, equality constraints, but these are easily dealt with using Lagrange multipliers $\eta$ and $\theta$ in order to remove the constraints by adding them to the Lagrangian 
$\mathcal{L}(w,\eta,\theta) = \frac{1}{2}w^\intercal \Sigma w + \eta(R^\intercal w-\mu)+\theta(A^\intercal w-b)$. 
In this case, the portfolio optimization has a closed form solution that can be obtained by solving a single linear system of dimension $n+2$:
\begin{align} \label{LS}
 \begin{pmatrix} 
\mathbf{0} & \mathbf{0} & R^\intercal \\
\mathbf{0} & \mathbf{0} & A^\intercal  \\
R & A & \Sigma \\
\end{pmatrix} 
\begin{pmatrix} 
\eta  \\
\theta   \\
w \\
\end{pmatrix} = 
\begin{pmatrix} 
\mu  \\
b   \\
\mathbf{0} \\
\end{pmatrix}
\end{align} 
We refer to this as the ``unconstrained'' portfolio optimization problem, and this is the simplest and least complex variant of the problem. 
This problem admits a simple polynomial time classical algorithm.
Nevertheless, it has been suggested that quantum computers might provide a speedup for this problem \cite{rebentrostlloyd2018quantumcf}, at least in certain quantum hardware scenarios.
We will discuss this approach in detail in Section \ref{subsec:unconstrainedportfolio}.

In many cases of practical interest, one needs to add more constraints in the portfolio optimization problem. 
The constrained portfolio optimization problem arises when one adds positivity constraints, these constraints are general enough to include linear budget constraints. 
The constrained portfolio optimization problem is given as:
\begin{align*}
\min _{w\in\R^n} & w^\intercal \Sigma w \\ 
\text { subject to } & R^\intercal w=\mu \\ 
& A^\intercal w= b \\ 
& w_{i} \geq 0 ~\text{ for all $i\in\{1,\dots,n\}$}
\end{align*}
The positivity constraints $w_{i} \geq 0$ indicate that the portfolio is long-only so the investor is not allowed to sell assets he does not own. 
More general inequality constraints $C^\intercal w \geq d$ can be included in this formulation by introducing the slack variables $s = Cw -d$ and adding the positivity constraint $s \geq 0$. 

Such linear budget constraints can encode complex requirements, for example that at least a certain fraction of the portfolio is invested in a certain asset class or that the budget for investing in certain asset classes is restricted. 
This constrained portfolio optimization problem is still a convex optimization problem and in fact it can be reduced to what is known as a second order cone program.
Therefore this problem does admit a polynomial time classical algorithm --- however this algorithm is substantially more challenging in practice than the unconstrained case.
Recent work of Kerenidis, Prakash and Szil\'{a}gyi \cite{kerenidis2019quantum} has shown a polynomial quantum speedup for this problem. 
We discuss quantum algorithms for this form of portfolio optimization in Section \ref{subsec:constrainedportfolio}.

There is a third type of constraints one can add to portfolio optimization, which are integer constraints, for example in order to limit the number of assets to invest in. Such a constrained portfolio optimization problem with integer constraints can be reformulated as a Mixed Integer Program (MIP)
\begin{align}\label{eq:MIP}
\min _{\substack{w\in\R^n\\ y\in\{0,1\}^n}} & w^\intercal \Sigma w \\ 
\text { subject to } & R^\intercal w=\mu \\ 
& A^\intercal w= b \\
& \sum_{i=1}^{n} y_{i} \leq K \\ 
& 0 \leq w_{i} \leq y_{i} 
\end{align}
Here, the $0/1$ integer variable $y_i$ indicates if there is investment in asset $i$ or not, and the constraint $\sum_{i=1}^{n} y_{i} \leq K$ limits the number of assets to invest in to at most $K$. A mixed-integer constrained optimization problem is a much harder computational problem, in fact it is NP-complete, and combinatorial optimization heuristic techniques can be used to find feasible good solutions. 
A number of quantum heuristic algorithms have been proposed to solve such combinatorial optimization problems, many of which are designed to be implementable in near-term quantum hardware.
We discuss quantum algorithms for this form of portfolio optimization in Section \ref{sec:portfolio_combinatorial} --- both general approaches to this problem through quantum annealing and circuit-based methods, as well as methods recently developed by QC Ware to improve upon these heuristics \cite{adame2020inhomogeneous}.

\subsection{Convex Optimization techniques for portfolio optimization}
\label{sec:portfolio_convex}

Here we describe how quantum computing might provide speedups for portfolio optimization in the case of convex constraints. We will start by looking at the unconstrained case and continue with the more realistic case where the positivity and budget constraints have been added. 

The main idea is to use the power of quantum computers to perform highly efficient linear algebra and apply it to both the unconstrained and constrained version by solving one linear system in the former case, and by performing an Interior Point method in the latter case.  

Broadly speaking, quantum computers can perform linear algebra efficiently by using a quantum state on $n$ qubits to encode a $N=2^n$ dimensional vector.
More specifically, since a system of $n$ qubits is described by a normalized complex vector of dimensionality $N=2^n$, one can think of a quantum state as representing a vector $x\in\mathbb{C}^{N}$ (up to its norm), and storing its norm separately.
If the desired linear algebra operation can be performed ``natively'' by the quantum computer, then by acting on this quantum state, one can produce a quantum state that \emph{encodes} the vector after this transformation is applied.
This approach was first taken by Harrow, Hassidim, and Lloyd \cite{HHL09}, who showed that one could solve linear systems efficiently in this model.
That is, given a matrix $M$ satisfying some conditions, if one is given a quantum state which encodes the vector $b$, then one can produce a quantum state encoding a vector $x$ such that $Mx=b$ in time which merely scales with the \emph{logarithm} of the dimension of $x$.
In contrast classical algorithms which solve $Mx=b$ all run in time which scales at least linearly (and usually polynomially) in the dimension of the vector $x$.
Therefore at first glance these algorithms seem to perform these linear algebra operations \emph{exponentially faster} than classical methods. 
These sorts of algorithms have been generalized in many subsequent works (e.g. \cite{GLSW18}).

While these methods can be said to offer exponential quantum speedups over classical method, in reality there are many caveats to these algorithms which must be taken into account and which could diminish the degree of quantum speedup (see e.g. \cite{childs2009equation, A15}).
For one, the \emph{output} of a quantum linear system solver solving $Mx=b$ is not a classical description of the vector $x$, but rather a quantum state \emph{encoding}\footnote{More formally, if $i$ is a basis for the vector space, then the quantum state is a vector whose $i$th amplitude is proportional to the $i$th entry of $x$.} the vector $x$. 
This sort of representation is sufficient in certain algorithmic scenarios but not others.
For example, if one only wishes to learn heavy elements of $x$, or a coarse approximation to it, then sampling from the quantum state encoding $x$ could suffice. However if one wishes to learn a complete description of $x$, this in general requires $\Omega(N)$ time, which removes the exponential speedup, while potentially still keeping a polynomial speedup. 
Second, one must have a method of loading the \emph{input} to the problem efficiently into a quantum state. 
It has been suggested that in the future one might be able to do this operation natively in hardware using a quantum form of RAM (called q-RAM \cite{GLM08}), but thus far this remains speculative \cite{A15} and therefore more of a long-term candidate for implementation if hardware progress allows it.
Instead recent work has focused on more direct circuit-based methods of loading the quantum data (such as quantum data loaders \cite{patent2}) that are amenable to more near-term or intermediate-term implementation. 
Additionally there are many parameters in the runtimes of these algorithms --- such as dependencies on the condition number of the matrix $A$, etc. --- that one must take carefully into account.
Finally, there are several classical algorithms whose runtimes also depend poylogarithmically on the dimension of the vectors in analogous input/output models, but with substantially worse dependencies on the condition number and other parameters \cite{gilyen2018quantum,gilyen2020improved}. Therefore the right way to describe these quantum speedups may not be to merely state them as exponential or polynomial, but to see for each specific application how these theoretical speedups translate in practice.

In any case, even taking into account these factors, in many cases one can still derive quantum algorithms for optimization problems using the quantum linear algebra approach with rigorous performance guarantees, and which improve polynomially on the best known classical algorithms.
In particular this has been shown for certain variants of convex portfolio optimization in recent work of Kerenidis, Prakash, and Szil\'{a}gyi \cite{kerenidis2019quantum}.
We will describe this algorithm shortly in more detail. 

In general these algorithms have medium to high requirements for the quantum hardware needed to run them --- and therefore are not strong candidates for near-term implementation in the next several years.
One major bottleneck is that the quantum algorithms for solving linear systems and other quantum linear algebra primitives all rely on quantum algorithms for Hamiltonian simulation --- a problem naturally defined in physics in chemistry which basically asks, given a Hermitian matrix $A$, to apply the matrix $e^{-iA}$ to a quantum state.
This is a fundamental ``quantum'' capability out of which other linear algebra primitives can be derived.
Unfortunately algorithms for Hamiltonian simulation are quite complex and difficult to implement on quantum hardware.
For instance recent work by Reiher et al. \cite{reiher2017elucidating} suggested these algorithms would require very low-error quantum devices to run these algorithms ($\ll 10^{-9}$ per two qubit operation).
Therefore neither Hamiltonian simulation nor quantum linear systems solvers are immediate candidates for near-term implementation on quantum devices, but rather longer-term prospects.

However, there is still a lot that can, and still needs to, be done to bring these algorithms closer to implementation, and we will describe some approaches shortly (see also Section 4).
First, in order to be able to solve the linear system $Mx=b$ fast on a quantum computer, one needs to be able to create a quantum state which encodes the vector $b$, and our work on data loaders \cite{patent2} can perform such a task with a logarithmic depth quantum circuit. Second, in approximate methods for solving linear systems, one performs a number of matrix multiplications rather than inversions, and these operations boil down to estimations of inner products between quantum states, another operation that can be efficiently performed. Third, the NISQ techniques for amplitude estimation we described in the previous section can potentially also be useful for reducing the resources necessary for the steps of amplitude amplification or even phase estimation, which are used during the quantum linear algebra procedures. 
Overall, while we cannot say right now that quantum linear algebra can be performed with NISQ machines, there are many tools that can be used to reduce considerably the necessary resources and we believe that the NISQification of such procedures is both extremely important and also a viable goal.

In the next sections we will describe how these quantum algorithms can be applied to the unconstrained and constrained variants of portfolio optimization.

\subsubsection{Quantum speedups for Unconstrained Portfolio Optimization}
\label{subsec:unconstrainedportfolio}

As we said before, the unconstrained portfolio optimization problem can be easily reduced to linear least squares problem, that can be solved with a single call to a quantum 
linear system solver. The linear system to be solved is of the form $Mx=b$ where the matrix $M$ is given in terms of the problem formulation as in Equation~\eqref{LS}.

Rebentrost and Lloyd used this reduction in \cite{rebentrostlloyd2018quantumcf} to provide a quantum algorithm for solving the unconstrained portfolio optimization. The efficiency of the quantum algorithm depends directly on the efficiency of the quantum linear system solver. Here, one needs to be very careful, since the running time of a quantum linear system solver depends on a number of parameters, including the dimension of the linear system, the condition number of the matrix $M$, the ability to have quantum access to the matrix $M$ and the vector $b$, the sparsity or the Frobenius norm of the matrix $M$, and the approximation error. 

More precisely, we have the following precise statement about the state of the art quantum linear system solvers  by \cite{CGJ18, GLSW18}. 
Let $A \in \R^{N\times N}$ be a matrix with non-zero eigenvalues in the interval $[-1, -1/\kappa] \cup [1/\kappa, 1]$, and let $\epsilon > 0$. 
Given that we can have quantum access to the matrix $A$ in time $T_{A}$ and can prepare the state $\ket{b}$ in time $T_{b}$, then a  state $\epsilon$-close to $\ket{A^{-1} b}$ can be generated in time $ \O((T_{A} \kappa \zeta + T_{b} \kappa) \log(\kappa \zeta /\epsilon))$, 
where $\zeta(A) = \min( \norm{A}_{F}/\norm{A}_{2}, \mbox{sparsity}(A))$.

Note as well, that the outcome of this procedure is a quantum state close to $A^{-1} b$ and not the classical solution $ A^{-1} b$. Obtaining an $\epsilon$-approximation to the classical solution involves running the quantum algorithm a number of times, in particular $ \O(N/\epsilon^2)$ times for an $\epsilon$-approximation in the $\ell_2$-norm and $ \O(1/\epsilon^2)$ times for an $\epsilon$-approximation in the $\ell_\infty$-norm, and sampling from the quantum solution. This factor, needless to say, must be added to the quantum running time as well. 

As previously mentioned, we emphasize that using quantum linear system solvers are not always ``exponentially" faster than the classical ones, as it is widely and wrongly stated, but a precise analysis is necessary in order to figure out in what cases the quantum linear system solver indeed offers a speedup, and what is the magnitude of this speedup. Deep expertise in quantum algorithms can result to 
much larger speedups than using an out-of-the-box quantum linear system solver, since different variants of the quantum algorithms can optimize the dependence on the parameters for the specific use case. 

We can now go back to the work of \cite{rebentrostlloyd2018quantumcf} and state more precisely that the quantum running time is $  \O(\frac{N}{\epsilon^2}(T_{A} \kappa \zeta + T_{b} \kappa) \log(\kappa \zeta /\epsilon))$ using the state-of-the-art quantum linear system solvers \cite{CGJ18, GLSW18}.

In the best case, most parameters are small, namely logarithmic in the time $T$ and assets $N$, and hence the quantum running time can become $ \tilde{\O}(N/\epsilon^2)$ to have a solution $\bar{w}$, which is $\epsilon$-close in $\ell_2$-norm to the optimal portfolio $w$.  This can in cases provide a considerable speedup since for the classical algorithm for portfolio optimization, it takes $\O(TN^2)$ operations to build the covariance matrix and $\O(N^3)$ operations to implement the pseudo-inverse to find the optimal portfolio. 

Here we do not make a more fine-grained analysis on the potential speedups (including number of qubits, circuit depth and error rates), since unlike quantum algorithms for Monte Carlo that have been heavily optimized for NISQ machines, there is still much theoretical work that needs to be done in order to find the most NISQ versions of quantum linear system solvers. For now, these algorithms remain longer term.

\subsubsection{Quantum speedups for Constrained Portfolio Optimization}
\label{subsec:constrainedportfolio}

We now describe how quantum computers may attain a speedup for solving constrained portfolio optimization.
In particular we will explain Kerenidis, Prakash and Szil\'{a}gyi's \cite{kerenidis2019quantum} recent quantum algorithm for the portfolio optimization problem with $r$ positivity constraints.
The algorithm works by reducing the problem to a second order cone program (SOCP) instead of quadratic program. 

The quantum algorithm is based on an interior point method that requires $ \O(\sqrt{r} \log (1/\epsilon) )$ iterations, where a linear system of a form similar to Equation~\eqref{LS} is solved in each iteration. The solutions to the linear systems at each iteration are recovered using a fast quantum tomography procedure \cite{KP18}, and these solutions are then used to write the linear system for the next iteration. The quantum interior point method is suitable for problems where the solutions need to be recovered to low accuracy, this is due to the use of the low precision quantum tomography subroutines unlike classical linear systems solvers that output exact solutions. Portfolio optimization is such a problem, where the desired accuracy is comparable to the amount of noise in the input data.

The quantum algorithm has a running time of 
\begin{align*}
\O\left(N \frac{\zeta \kappa}{\delta^{2}} \sqrt{r} \log (1 / \epsilon)\right)
\end{align*}
where the parameters are as follows: 
\begin{itemize}
	\item $r$: the number of positive constraints, which is called the rank of the SOCP. The unconstrained portfolio optimization corresponds to the case $r=1$. 
	\item $\epsilon$: the duality gap for the solutions to the SOCP. 
	\item $\delta, \zeta$: problem-dependent parameters quantifying the distance of the intermediate solutions from the cone-boundary. 
	\item $\kappa$: problem-dependent parameter, equal to the maximum condition number of the intermediate matrices in the interior point method for the SOCP. 
\end{itemize}
The best known classical algorithm for this problem has running time 
\begin{align*}
\O\left(N^3 \sqrt{r} \log (1 / \epsilon)\right).
\end{align*}
The quantum algorithm achieves a speedup over the classical algorithm in a regime where $ \frac{\zeta \kappa}{\delta^{2}} < N^{2}$. 
The problem dependent parameters need to be estimated for real world instances in order to validate the claims for a quantum speedup. 

An experimental evaluation of constrained portfolio optimization with positivity constraints was carried out in \cite{kerenidis2019quantum} for subsets of the cvxportfolio dataset consisting of daily stock prices for the S\&P-500 companies. The results indicate that the quantum algorithm can achieve a substantial polynomial speedup over the classical 
algorithm with an estimated running time exponent of $ \O(N^{2.37})$ for this dataset. We expect similar polynomial speedups for practical portfolio optimization instances, as the second order cone programs would only need to be solved to moderate accuracy for portfolio optimization over noisy real world datasets. 

Again, the main limitation of the algorithm is its dependence on quantum linear system solvers \cite{CGJ18, GLSW18} that require large depth circuits and remain beyond the reach of 
near and intermediate term devices. 
We remark, nevertheless, that the quantum resources for solving this type of constrained portfolio optimization are the same as in the unconstrained case, since we used a hybrid algorithm, where a classical interior point method was used on top of a quantum linear system solver. Algorithmic improvements that lower the depth of quantum linear system solvers can bring quantum constrained portfolio optimization closer to implementation.

\subsection{Combinatorial Optimization techniques for Portfolio Optimization}
\label{sec:portfolio_combinatorial}

Many quantum algorithms have been proposed to address the integer constraint portfolio optimization problem using combinatorial optimization techniques.
In its most general formulation, constrained portfolio optimization as described in Equation~\eqref{eq:MIP} is a mixed integer program (MIP), which is known to be NP-hard \cite{gareyjohnson1979, mansini1999heuristic}.
Under widely believed computational assumptions, this means there is no efficient classical or quantum algorithm that can solve all instances of these problems.
However, this does not mean one cannot solve this problem in practice, for several reasons.
First, usually one only wishes to solve particular instances of the portfolio optimization problem, namely those which naturally arise in the market. These instances could be easier to solve than worst-case instances.
Second, oftentimes one is satisfied with finding approximately optimal solutions to the portfolio optimization problem, rather than the true optimal solutions.

Therefore, in practice a number of heuristic algorithms have been proposed to solve integer constrained portfolio optimization.
On the classical side, a large number of heuristics have been proposed, both for the specific problem of portfolio optimization (e.g. \cite{speranza1996heuristic,mansini1999heuristic,bertsimas1999portfolio,cura2009particle,angelelli2012kernel}) and for the more general problem of mixed integer programming (e.g. \cite{cote1984large,balas2004pivot,pochet2006production}) by the operations research community. 
We do not survey these classical methods here, but refer the reader to the MIP and portfolio optimization literature.

A number of \emph{quantum} heuristic algorithms have been proposed for portfolio optimization as well.
Here we will discuss two categories of such quantum algorithms: quantum annealing based approaches and circuit model based-approaches.

\subsubsection{Quantum annealing based approaches}
In quantum annealing heuristics, the basic idea is to encode the solution to a portfolio optimization problem as the physical ground state (lowest energy state) of a quantum system \cite{apolloni-etal-1989,kadowaki-nishimori01998,finnila-etal-1994,farhi-etal-2000}.
While in general quantum systems do equilibrate to their ground state over time, in the worst case the time it takes to reach the ground state can scale exponentially in the size of the system.
Which annealing times are necessary and sufficient for which problem instances is a major research question, both theoretically and experimentally. 

To improve the probability that the system settles into its ground state, a well studied algorithm is that of quantum adiabatic optimization \cite{farhi-etal-2000}, which is a specific approach to quantum annealing. 
In the adiabatic quantum computation, one initially prepares a ground state of a very simple system, then slowly changes the system parameters to those of the target system whose ground state encodes the solution to the optimization problem.
If this transition from the initial system to the final system occurs slowly enough, then the final state will indeed be close to the intended ground state solution. 
A sufficient condition for what is ``slow enough'' is given by the quantum adiabatic theorem, which requires an analysis of the spectral gap of the quantum systems involved. 

Quantum adiabatic optimization is in many ways a quantum analogue to the classical simulated annealing algorithm.
First, there is a similar intuition that the annealing process allows the system to settle into the ground state without being trapped in local optima.
While classically the probability of escaping from a local optimum is controlled by the thermal fluctuations of the annealing algorithm (which controls the probability of jumping to less desirable solution), quantum systems can escape local optima through quantum fluctuations; a process also known as quantum tunneling.
Quantum tunneling is known to be able to overcome barriers that are impenetrable classically, hence the promise is that quantum annealing devices may be able to outperform classical devices for optimization problems. 
The extent to which quantum annealing can live up to this promise has been studied extensively \cite{farhi2002quantumadiabaticvsSA, denchev2016computational, Brady-van_Dam-2016}, however many open questions remain about this promising heuristic.

Also, similarly to classical annealing, oftentimes the theoretical convergence guarantees are quite slow (based on the spectral gaps of certain operators), but in practice one runs the algorithms for much shorter timescales in the hopes of finding a good enough solution quickly.

In general the power of this model is equivalent to that of standard quantum computation \cite{aharonov1999quantum}, and therefore offers exponential speedups over classical simulated annealing for certain problems.
However existing hardware implementations of quantum annealing are both highly noisy and only implement a small range of controllable system parameters which place them in a special subclass of annealing known as ``stoquastic'' annealing. Only recently has it been discovered that stoquastic annealing algorithms can potentially exponentially outperform classical computing, albeit so far only for a toy problem designed to give stoquastic annealing an advantage, and in a noiseless setting \cite{hastings2020power}.

A number of companies offers specialized quantum annealing hardware, most notably D-Wave systems.
These experimental devices implement a highly noisy form of stoquastic quantum annealing on up to a few thousand qubits, allowing for the implementation of many annealing algorithms in hardware. 
Although this device has much more noise than the idealized form of annealing mentioned above ---  as their device ``decoheres'' many times during the course of the annealing, which should not occur in an ideal annealing device --- their devices have demonstrated some forms of quantum effects. 
There has been a healthy debate in the academic community as to what extent quantum effects in the D-Wave device demonstrate speedups over classical computing (see e.g. \cite{shin2014quantum,albash2015reexamining}). 
Thus far the weight of the evidence has not supported a quantum speedup from existing hardware, though comparing the performance of heuristic algorithms vs. the DWave annealer is quite subtle, as one wishes to compare the D-Wave device to ``the best'' classical heuristic solver, and the latter term is not well-defined in the heuristic context.
That being said, as D-Wave is one of the few quantum devices of sufficient size to test quantum algorithms, there has been considerable interest in experimenting with the device to evaluate its performance.

The D-Wave device only works to solve optimization problems in a particular form known as a Quadratic Unconstrained Binary Optimization problem, or QUBO. In this problem one is given a matrix $A\in\mathbb{R}^{n\times n}$, and the goal is to find the minimizing bit string
\begin{align*}
\argmin_{x\in \{0,1\}^n} x^\intercal A x 
\end{align*}
Here the number $n$ corresponds to the number of qubits in the device.
The D-Wave device has a number of restrictions on exactly which constraint matrices $A$ can be implemented in its hardware.
Most prominent of these is a geometrical restriction --- the qubits of their device do not have all to all connectivity, but rather their interactions are limited to a particular graph known as the Chimera graph.
The non-zero entries of $A$ are required to lie on the edges of this graph.
It is known that solving QUBOs on this restricted graph remains NP-hard --- so in principle essentially any optimization problem can be mapped to a QUBO on the chimera graph with polynomial overhead.

However, this does mean that to apply the D-Wave device to a portfolio optimization problem of practical interest, one must first formulate the problem as a QUBO, and then embed the QUBO in the geometrical constraints of the Chimera graph, before running the quantum annealing algorithm in hardware. This is known as the minor embedding problem, as one often performs this by identifying several of the qubits with one another, and thereby embedding the problem of interest as a minor of the Chimera graph \cite{choi2008minor,choi2011minor}.
Finding an efficient mapping for this problem of interest is one of the key challenges of using the D-Wave device in practice, as one often finds that the overheads required for the embedding come at a high cost in the number of qubits. 
For many optimization problems one can only solve instances with hundreds of bits, even though the device has thousands of bits, due to these geometrical constraints (see e.g. \cite{rieffel2015case}).

For the portfolio optimization problem, this mapping is particularly challenging as the DWave hardware naturally encodes problems over binary variables, but portfolio optimization is most naturally expressed with integer variables. Therefore, in addition to the overheads mentioned above, there are additional overheads in the embedding due to encoding these integer values either in binary or in unary (simpler, but with more overheads).
Moreover, certain constraints in the MIP are ``hard constraints'' (they must be satisfied) while others are ``soft'' (merely penalized in the objective function). Annealing hardware naturally only has the ability to impose soft constraints. In theory a soft constraint can be made ``hard'' by increasing its penalty in the objective function. However in practice there are limitations in the range of constraints one can impose, so one is forced to find workarounds. 
As a result, existing experimental implementations in D-Wave hardware have been limited in the size of the problems they can implement in hardware.

A number of prior works have considered using D-Wave annealers for portfolio optimization.
Prado \cite{prado2015trajectories} and Marzec \cite{marzec2016quantum} suggested portfolio optimization as a potential application of quantum annealers. 
Subsequently Rosenberg et al.\ \cite{rosenberg2016quantumannealer} described several different schemes for embedding portfolio optimization as a QUBO, and ran this annealing algorithm on the 512-qubit D-Wave annealer. Due to embedding overheads they were only able to run very small instances in hardware (e.g. 2 or 3 assets, 2 or 3 time periods, and maximum holdings of at most 3 in each assets). 
Elsokkary \cite{elsokkary2017financial} similarly ran small portfolio optimization problems on the D-Wave annealer.
More recently Ventuerlli and Kondratyev \cite{venturelli2019reverse} described an alternative annealing approach to portfolio optimization using ``reverse annealing." In this variant of annealing, one begins with a (non-optimal) classical solution to the optimization problem (say found by a classical heuristic), and then tunes the system towards a more ``mixing'' setting and then back again to the original problem. The hope is that this additional mixing might help to ``jostle'' the solution of the classical local optimum via quantum tunneling. 
They run this algorithm on with up to 60 assets on a 2048-qubit annealer, and found significant improvements over traditional quantum annealing methods, and also discuss mechanisms for embedding the problem in QUBO form. 
Very recently, Grant et al. \cite{grant2020} considered several ways of embedding portfolio optimization onto the D-Wave hardware and compared different annealing schedules on each, and Cohen et al.\ \cite{cohen2020portfolio} ran a 40-qubit portfolio optimization problem on the D-Wave annealer as well.
Mugel et al.\ \cite{mugel2020dynamic} ran larger scale portfolio optimization problems on D-Wave using D-Wave's hybrid solver, using a combination of pre-processing and D-Wave's hybrid solver which breaks the problem into smaller pieces to run 65 fully connected qubits at a time (the largest size that fits on their 2048 qubit device). 
Although the D-Wave approach found a solution more quickly, they found mixed results in terms of solution quality compared to a quantum-inspired classical algorithm.

Together these works have provided proof of principle demonstrations that portfolio optimization problems can be mapped to existing quantum annealing hardware, albeit with significant overheads due to a) geometrical constraints on interactions and b) the encoding of integer variables into binary variables. It remains to be seen if these heuristic quantum solvers will outperform classical heuristics, as we have not yet had the quantum hardware to implement these problems at scale. We discuss several future directions and open problems in this area in Section \ref{sec:portfolio_combinatorial_futuresprospects}. 

Additionally, we note that it is a very active area of research to improve the performance of these D-Wave based optimization heuristics.
In addition to the ``reverse annealing'' approach mentioned previously \cite{venturelli2019reverse}, another method that has demonstrated improvements over ``vanilla'' quantum annealing approaches is the anneal offsets approach of Adame and McMahon \cite{adame2020inhomogeneous}. 
Here the idea is relatively simple.
In standard quantum annealing, the penalty terms for violating a constant of the QUBO are all turned on homogeneously at the same rate.
That it, all variables in the QUBO are treated equally.
However in many practical optimization problems, some of the variables tend to ``freeze out'' early in the run of the adiabatic algorithm, i.e. the values of these variables become essentially classically fixed midway through the quantum anneal. 
This may be undesirable as it limits the extent to which the device is exploring the parameter space quantumly.
The basic idea behind anneal offsets is to apply the penalty terms to those variables \emph{more slowly} than those of the other variables.
In other words, by driving the system \emph{inhomogeneously} towards the ground state, one might be able to improve the performance of the annealer.
Several schemes have been proposed to perform this sort of inhomogeneous driving of the annealing schedule in order to improve the performance of quantum annealing (see e.g. \cite{lanting2017experimental}).
However it might be unclear \emph{a priori} which variables might tend to ``freeze out'' in a QUBO, and therefore might be difficult to design this inhomogeneous driving schedule. 
Adame and McMahon \cite{adame2020inhomogeneous} described a particularly simple heuristic algorithm for choosing the inhomogeneous driving schedule, and empirically showed this can provide orders of magnitude improvement in time to solution compared to standard quantum annealing, particularly on problem instances which the annealer found difficult.
This result suggests that there is still much to optimize in the implementations of these quantum annealing heuristics before comparing them to classical solvers.

\subsubsection{Quantum circuit model based approaches}
A second family of approaches for portfolio optimization are based on universal circuit based machines, rather than special-purpose annealing hardware. 
Generally speaking these approaches fall into two categories: near-term approaches based on the Quantum Approximate Optimization Algorithm (QAOA) \cite{farhi2014quantum} and its variants \cite{hadfield2019quantum}, and longer term approaches based on amplitude amplification.

The QAOA \cite{farhi2014quantum} is a quantum algorithm which seeks to find approximate solutions to optimization problems.
That is, given a set of $m$ constraints over binary variables, the goal is to find a solution which satisfies \emph{most} of the constraints. 
This can be a far easier task than finding an exact solution: for example, many NP-hard problems have efficient approximation algorithms \cite{vazirani2013approximation}. 
The goal of the QAOA is to similarly find an approximate solution using a particularly simple form of quantum circuit.
It describes an infinite family of algorithms, one for each integer $p$, of increasing circuit depth. 
While it is known that very high-depth versions of the algorithm do admit speedups over classical algorithms \cite{lloyd2018quantum}, much of the focus of the literature has been on low-depth (say $p=1$ or $2$, or a small constant) implementations of the algorithm, because these versions are most amenable to running on near-term quantum devices without error correction.
It is not known if the algorithm offers a computational speedup over classical methods at these low depths; to date its worst-case performance guarantees have been eclipsed by classical optimization algorithms based on SDPs (e.g. \cite{goemans1995improved}), or in some cases even simple classical heuristics \cite{hastings2019classical}. 
Moreover recent work has suggested that for particular problems, QAOA may need to have super-constant depth to match classical algorithms' performance \cite{farhi2020quantum}.
Nevertheless it remains an open area of investigation if the algorithm could be useful as a heuristic classical solver.

A few prior works have applied the QAOA framework to portfolio optimization, although overall the area remains underexplored.
For example, Hodson et al. \cite{hodson2019portfolio} formulated implementations of QAOA and its variants for portfolio optimization algorithm and tested it on small instances (8 assets) in simulation. Barkoutsos et al.\ implemented QAOA for a small portfolio optimization problem (with 6 qubits) on the IBM Poughkeepsie processor, as part of a broader work testing variants of QAOA objective functions. 
Mugel et al.\ also ran a QAOA-related algorithm\footnote{More specifically, they ran a more general algorithm known as VQE and a variant they called VQE-Constrained.} to solve portfolio optimization in simulation, but were similarly constrained to run very small problem instances.

These studies have focused on relatively small portfolio optimization problems because of hardware limitations.
In general the number of qubits available on circuit-based architectures is much smaller than on annealing-based architectures.
Additionally, the circuit depth one can apply is very limited by high per-gate noise rates.
Even in simulation, the number of qubits one can simulate (via direct simulation of the wavefunction) is limited to ~30 due to exponential number of parameters needed to describe the wavefunction. 
This strongly limits the size of the portfolio optimization problem that can be explored with QAOA.

Finally we briefly mention a final class of algorithms for combinatorial optimization, which is algorithms based on amplitude amplification. 
Large-scale fault tolerant quantum computers will be capable of quadratic speedups for many classical optimization algorithms.
For instance, Grover's search algorithm can be used to search a solution space of $2^n$ items in merely $2^{n/2}$ time. 
This algorithm can be generalized to provide speedups for many classical heuristic optimization algorithms as well, such as branch and bound methods \cite{montanaro2020quantum}.
There are also related algorithms to provide quantum speedups for classical backtracking algorithms via quantum walks \cite{montanaro2018backtracking}.
Generally speaking the hardware requirements for these methods are high, and comparable to those required to run the standard quantum Monte Carlo algorithm (see Section \ref{sec:MCresourceestimates} as well as \cite{campbell2019applying} for resource estimates). 
Therefore it is unlikely these algorithms will be implemented in near-term hardware, but are an interesting subject of theoretical evaluation.
These approaches could have long-term impacts similar to those for quantum Monte Carlo speedups.

\subsubsection{Prospects for combinatorial optimization}
\label{sec:portfolio_combinatorial_futuresprospects}

There are many important directions to explore regarding the potential impact of quantum combinatorial optimization algorithms on finance.

First, it remains open whether or not near-term amenable quantum heuristics (annealing and QAOA) offer a speedup over classical methods. 
These algorithms were designed to be implementable in near-term hardware, but we do not have evidence that they outperform classical methods.
The major open question in this area is whether or not these offer bona fide speedups, or if they can be beaten by classical methods.
While there is much theoretical work to be explored here --- for example, does QAOA or annealing offer worst-case performance improvements over classical analogues? \cite{hastings2019classical} --- the most relevant practical question is whether or not these provide speedups for typical instances faced in practice.
This will require a careful comparison with best-in-class classical heuristic solvers (rather than ``straw man'' comparison with brute force methods, or solvers not tailored to the application), as well as with classical approximate solution methods \cite{charikar2004maximizing,alon2005quadratic,hermelin2020}.
This is a tricky problem to address, as one may not have access to the classical heuristics used by market players as they may contain proprietary information. 

As a result, we call for the creation of a public collection of portfolio optimization heuristics with which we can establish clear benchmarks for the performance of quantum algorithms.
These would not necessarily be the tuned heuristics used in practice (which may contain proprietary information), but could be related algorithms with similar methodologies.
This collection would establish a baseline for classical heuristics to be compared with annealing or QAOA approaches.

Second, further work is needed to optimize the embedding of portfolio optimization algorithms into the constraints of hardware.
Several aspects of this problem do not naturally map to the quantum binary optimization setup --- the fact the variables are naturally integer rather than binary, the fact the optimization problem tends to have all-to-all couplings (as every asset is correlated with every other one), and the fact that certain constraints are ``hard constraints'' that must be satisfied (and not just approximately so). Each of these restrictions presents a major challenge for solving production-scale instances of portfolio optimization (or more generally, mixed integer programs) in real quantum hardware, and further work is needed to explore optimal embeddings.

Third, further work is needed to understand proveable speedups for portfolio optimization problems. For instance, can amplitude estimation techniques provide quantum speedups for other classical portfolio optimization heuristics such as pivot and shift? There are many interesting quantum algorithmic questions to be answered here which may have medium to long term impacts in the area.

Fourth, further work is needed to estimate the resources needed to run these algorithms at scale, and to determine the point at which point they offer an advantage over classical methods. 
There are concrete factors which are preventing the implementation of these algorithms at sufficient scale so that they can be compared to classical heuristics for interesting problem sizes.
On the annealing side, the limitation is in the \emph{quantity} and connectivity of the qubits. 
The QUBO formulation of portfolio optimization usually has all to all constraints, but the limited connectivity of the existing hardware means that they can only run problem instances of much smaller sizes.
For example to date the largest quantum annealer to date, the D-Wave Advantage with 5000 physical qubits, can only run all-to-all connected QUBOS on up to 180 qubits \cite{boothby2020next} (though this will in practice be more limited due to hardware constraints and imperfections).
For estimates of how many all-to-all coupled qubits are needed to tackle portfolio optimization in the annealing setting with various embedding schemes, see \cite{rosenberg2016quantumannealer}.
Improving the quantity of connectivity of these annealing qubits would allow for larger problems to be run in hardware.
On the QAOA side, the \emph{quality} of the qubits is the major limiting factor. Although 50+ qubit machines are already available, which could in principle be used to encode all-to-all connected 50+ qubit optimization problems (by swapping qubits), their high two-qubit gate error rates prevent this from being a viable strategy.
More generally, these high error rates prevent the implementation of QAOA or related heuristics at higher $p$-values and higher depths, and on problems whose connectivity is not closely related to the underlying 2D qubit architectures. 
Therefore progress on qubit quality will be key to implementing QAOA on significant-sized problems.
Once these problems can be run at scale in hardware, the next question will be to determine the point at which these achieve a practical speedup. 
This is quite difficult to answer as we to not yet know if, and to what extent, these algorithms provide quantum speedups. 
It is clear, however, that much more progress is needed in hardware (both number of qubits, connectivity, and error rates) before production-scale problems can be implemented in quantum hardware, and the question of empirical/heuristic speedups can be evaluated.

\section{Machine Learning in Finance} \label{sec:ML} 
Machine learning has become increasingly important in recent times with the growth in the types and size of data sets and the development of high-performance machine learning algorithms for a wide range of data analysis tasks. It has been successfully applied to achieve state of the art performance for a variety of tasks in many domains, and the applications of machine learning to finance have recently started becoming more prevalent. For example, machine learning can be applied to automated portfolio management, algorithmic trading, risk assessment, fraud detection etc. For a deeper analysis of the applications of Machine Learning in finance we recommend \cite{jpmorgan17, vL17, fsb17}.
Here, we provide a brief description of how quantum computing can enhance machine learning applications, refraining at this time from a comprehensive analysis of quantum machine learning. 

Present day machine learning can be broadly sub-divided into supervised, unsupervised and reinforcement learning. 
Some fundamental machine learning tasks include regression and classification for supervised learning, clustering, dimensionality reduction and learning mixture models for unsupervised learning and learning Markov decision processes for reinforcement learning. For some of these tasks it is common nowadays to use neural networks (deep learning) instead of the more algorithmic machine learning, especially when one has large amounts of unstructured data, as in the case of image or text classification or playing video games. Machine learning is a very active area where new techniques are constantly being developed and the algorithms are improving in performance and applicability.

Machine learning algorithms work usually with real valued data and make extensive use of linear algebra sub-routines. The development of the field of quantum machine learning has been spurred by the development of quantum linear algebra algorithms \cite{HHL09} and more recent variants \cite{GLSW18}, which provide the possibility of exponentially faster linear algebra on quantum computers for certain types of matrices. However, a direct application of quantum linear system algorithms is known to have several caveats \cite{A15}, including the important issues of loading classical data into quantum computers and reading out classical outputs from quantum computations, as well as performance guarantees that are instance-dependent and not always easy to grasp. 

Most quantum machine algorithms working with classical data assume the availability of a quantum random access memory (QRAM), which are quantum analogs of classical Random Access Memory (RAMs) and are not expected to be physically realizable in the near future \cite{GLM08} . Recently, an alternate approach for encoding classical data into quantum states using parametrized circuits has been developed \cite{patent2} using data loaders. The data loaders are parametrized circuits that for encoding a vector $x \in \mathbb{R}^{n}$ dimensional vector into a quantum state $\ket{x}$, optimally require that the number of qubits $Q$ used and the depth $D$ of the loader circuit satisfy $QD = \O(n \log n)$. 
The parallel loader with $Q = n$ and $D= \log n$ and the matrix loader with $Q=2\sqrt{n}$ and $D= \sqrt{n} \log n$ are particularly useful for the development of near and intermediate term quantum machine learning algorithms. 

At the present time, no end-to-end application of quantum machine learning that achieves an exponential speedup over a classical machine learning algorithm is known, and several promising candidates for such speedups had to be reconsidered \cite{T18}. However, quantum machine learning remains a promising research area as large polynomial speedups over classical ML algorithms are relevant for many use cases, including those that arise in mathematical finance. In fact, in many cases and for instance sizes of interest, quantum machine learning algorithms can in theory provide a speedup factor of more than a million, compared to both classical and quantum-inspired algorithms \cite{KP16}.

In the following sections, we will discuss such quantum machine learning algorithms with potentially significant polynomial speedups for many fundamental learning tasks, with a view towards applications to finance. Given the bottlenecks for encoding classical data into quantum states and reading out the output of quantum computations, the speedups for quantum machine learning are not in the data gathering or cleanup phases of classical data analysis. Quantum machine learning algorithms can offer significant speedups on the data-processing part where the analogous classical algorithms perform computationally expensive linear algebra operations. Last, we will briefly discuss quantum neural networks and their applicability. 

\subsection{Supervised Learning}
Supervised learning corresponds to a setting where the input is a collection of labelled examples $(x, y)$ and the task is to predict the labels from the examples. Regression corresponds to the setting where the labels are real valued while classification corresponds to the setting where the labels are discrete valued. 
\subsubsection{Regression}
We discuss the quantum algorithms for linear regression/least squares and for k-nearest neighbors regression, these are quantum analogs of two of the most useful classical algorithms for regression. 

 Linear Regression or least squares is one of the most widely used machine learning algorithm for regression. The least squares problem has a closed form solution and is solved classically using linear system solvers for moderately sized data sets and an iterative gradient descent method for large datasets. Quantum methods corresponding to 
 both direct and iterative algorithms for least squares have been developed \cite{KP17, CGJ18}. These quantum algorithms remain intermediate term due to the use of high depth 
circuits for linear algebra. Among finance applications, linear regression is used for asset pricing and for the computation of multi-asset trend following strategies. 

 $k$-Nearest Neighbors is a non-parametric regression method that predicts the label for data point $x$ as the average of the $k$ labels of the data points closest to $x$ in Euclidean distance. It can be expensive classically due to the high dimensionality of data and the need to compute multiple Euclidean distances. Quantum k-nearest neighbor regression algorithms are more promising for nearer term applications as there are efficient quantum sub-routines for distance estimation and one can achieve a quadratic speedup compared to classical distance estimation. In fact, such efficient quantum distance estimators can be built by composing two of the data loader circuits followed by measurements \cite{patent2}. Among finance use cases, cross-asset risk premia investing is a possible application of k-nearest neighbor regression.

\subsubsection{Classification}
Classification is the task of predicting discrete valued labels from data. We discuss briefly quantum classification algorithms based on nearest centroids, support vector machines and convolutional neural networks. 

The Nearest Centroid classifier has as input a list of centroids and assigns a data point the label of the centroid closest to it. It is suitable for nearer term quantum implementations due to the availability of efficient distance estimation procedures and of clustering algorithms for finding the centroids. The nearest centroids classifier can be used in combination with unsupervised learning methods to extract more meaningful features before performing the classification.

Support vector machines are widely used for classification tasks and are the most accurate classical ML classifiers that are not based on neural networks. The support vector machines 
problem with $\ell_{2}$ norm penalties reduces to a single linear system and was proposed as a quantum algorithm by Lloyd, Mohseni and Rebentrost \cite{LMR13}. More recently, the quantum interior point algorithm for Second Order Cone programs was used to give a quantum algorithm for the SVM problem with $\ell_{1}$ penalties in the objective function \cite{KPS19a}. 

This is similar to to the portfolio optimization problem discussed earlier, where solving a single linear system gives an unconstrained/$\ell_{2}$ version of the problem while the quantum interior point method allows for the solution of the constrained/$\ell_{1}$ version of the problem. Experimental results indicate that the quantum algorithm achieves a polynomial speedup for the $\ell_{1}$-SVM, however the algorithm uses high depth quantum linear algebra sub-routines and is therefore not suitable for near term devices. Estimating trading currency volatility is a possible use case for quantum SVMs in finance.

\subsection{Unsupervised Learning}
Unsupervised machine learning corresponds to a setting where the input is an unlabeled dataset and the task is to build useful models from the unlabeled data. The prototypical tasks for unsupervised learning are clustering, feature extraction, learning mixture models and generative models. Quantum algorithms have been developed for many 
of these fundamental tasks, of which we first discuss clustering. 

\subsubsection{Clustering} 
The classical $k$-means algorithm is one of the most widely applicable clustering algorithms for real valued data. A quantum version $q$-means with the same guarantees as a robust version of the classical $k$-means algorithm and running time that is poly-logarithmic in data-set size has been proposed recently \cite{qmeans}. The qmeans algorithm is amenable to nearer term implementations as it can be implemented using distance estimation and quantum matrix multiplication. A possible finance use case for the qmeans algorithm is to portfolio risk analysis. 

The classical $k$-means algorithm generalizes to the Expectation Maximization algorithm for learning Gaussian mixture models, and the qmeans algorithm has also been generalized to 
quantum expectation maximization \cite{KLP20}, with a speedup similar to that for the qmeans algorithm. A quantum spectral clustering algorithm for data represented as a graph 
has been recently proposed \cite{KL20}, adding to the repertoire of near/intermediate term quantum clustering algorithms. Identification of market regimes (high/low volatility,rising/falling rates, rising/falling inflation) is a possible application for quantum expectation maximization in finance.

\subsubsection{Feature Extraction}
Projecting onto the eigenspaces of the data matrix is one of the most useful feature extraction or dimensionality reduction techniques in classical machine learning.  
Algorithms like principal component analysis (PCA) and linear discriminant analysis (LDA) utilize such projections. Quantum algorithms for principal component 
analysis \cite{LMR14} provide a speedup in terms of dimension but incurs a large overhead on the accuracy. More recent quantum projection algorithms use singular value estimation \cite{KP16} or more generally singular value transformation \cite{GLSW18}. 

Another projection/feature extraction technique is Slow feature analysis (SFA) which corresponds to projecting onto the low eigenspaces of a certain matrix. Quantum 
algorithms for slow feature analysis have been proposed recently \cite{KL18}, the method is also closely related to Linear Discriminant analysis (LDA). Other classical feature 
extraction methods utilize neural networks and auto-encoders, we next describe some recent works on quantum neural networks.

\subsection{Deep Learning}
Many approaches have been proposed for quantum deep learning based on parametrized circuits, we discuss here some of the recent algorithms that provably have performance comparable to the corresponding classical neural networks while offering quantum speedups for training. Quantum feed-forward neural networks in this paradigm were proposed 
in \cite{A18}, these can achieve significant savings in both the forward evaluation and backward training passes. More recently, a quantum convolutional neural network with speedups in training and back-propagation has been given \cite{KLP19}.  Quantum architectures for deep learning are an active research area, some important recent developments include an approach to quantum convolutional neural networks using variational circuits \cite{C19} and the development of quantum generative adversarial networks \cite{C19q}. 

Applications of deep learning to finance is an active research area suggesting that it also may be a promising area for quantum deep learning algorithms. Recently, deep learning based solutions have been found to significantly outperform classical approaches for finance applications like asset pricing \cite{chen2019deep}, mortgage risk \cite{giesecke2016deep} and limit order books \cite{sirignano2019deep}. However, deep learning tools are rarely used for finance applications like time series analysis, where tools from classical Machine Learning dominate \cite{jpmorgan17}.

\subsection{Reinforcement Learning}
Reinforcement learning corresponds to the setting where the machine learning algorithm interacts with an environment and takes actions that change the state of the environment with a view of optimizing a reward function. Recent work \cite{ karpe2020multi, daberius2019deep, ganesh2019reinforcement, deng2016deep} suggests that algorithmic trading is a promising 
application of reinforcement learning to finance. Quantum reinforcement learning is a relatively recent research area, finding an end-to-end application of quantum reinforcement learning to finance with significant speedups over classical algorithms remains a direction for future work.

\section{Conclusions}
Quantum computers are expected to have substantial impact on the Finance industry, as they will eventually be able to solve certain problems considerably faster than the best known classical algorithms. This is the case for example for some fundamental problems in Finance, including Monte Carlo methods, portfolio optimization and machine learning. However, it is a major challenge to determine when this impact will occur for each application, and, in fact, one of the most pressing challenges is to redesign quantum algorithms in order to both considerably reduce the quantum hardware requirements and at the same time keep their provable impact. In this work, we provided a review of the state-of-the-art, and we described several ways we have developed for designing or redesigning quantum algorithms for the NISQ era while maintaining performance guarantees. Finding the first quantum finance applications remains a challenge but advances in quantum algorithms together with better quantum hardware continue to bring these applications closer to reality.

\acknowledgments

We thank Kay Giesecke, Rajiv Krishnakumar, Ashley Montanaro, Nikitas Stamatopoulos, and Will Zeng for helpful discussions and comments on this manuscript.

\newcommand{\etalchar}[1]{$^{#1}$}

\end{document}